\documentclass[a4paper,11pt]{article}
\usepackage[margin=0.5in]{geometry}
\usepackage{enumerate}   
\usepackage{jcappub}
\usepackage{aas_macros}

\usepackage{amsmath}
\usepackage{amssymb}
\usepackage{bm}
\usepackage{natbib}
\usepackage{xspace}
\usepackage{lipsum}
\usepackage{hyperref}
\hypersetup{
	unicode,
	pdfauthor={Author One, Author Two, Author Three},
	pdftitle={Robust cosmology from cosmic shear with small-scale decontamination},
	pdfsubject={Robust cosmology from cosmic shear with small-scale decontamination},
	pdfproducer={LaTeX},
	pdfcreator={pdflatex}
}
\usepackage{graphicx, color}
\usepackage[utf8]{inputenc}
\usepackage[T1]{fontenc}
\usepackage{caption}
\usepackage{verbatim}
\usepackage[dvipsnames]{xcolor}
\usepackage{adjustbox}
\newcommand{\planck}{{\sl Planck}\xspace}
\newcommand{\tmpz}{2MPZ\xspace}
\newcommand{\wisc}{WISE$\times$SuperCOSMOS\xspace}
\newcommand{\bemu}{{\tt baccoemu}\xspace}

\newcommand{\fsky}{f_{\rm sky}}
\newcommand{\hpx}{\texttt{HEALPix}\xspace}

\newcommand{\nv}{\hat{\bf n}}
\newcommand{\enquote}[1]{``#1''}
\newcommand{\ihMpc}{ h\,{\rm Mpc}^{-1}}

\title{\boldmath Low-redshift constraints on structure growth    from CMB lensing tomography}
\author[a,b,c]{Andrea Rubiola,}
\author[d]{Matteo Zennaro,}
\author[d]{Carlos Garc\'ia-Garc\'ia,}
\author[d]{David Alonso}
\author[e,f]{{Raul E. Angulo}}

\affiliation[a]{Dipartimento di Fisica, Università degli Studi di Trento, via Sommarive, 14, 38123 Trento, Italy}
 \affiliation[b]{Dipartimento di Fisica, Università degli Studi di Torino, via P. Giuria 1, 10125 Torino, Italy}
\affiliation[c]{INFN – Istituto Nazionale di Fisica Nucleare, Sezione di Torino, via P. Giuria 1, 10125 Torino, Italy}
\affiliation[d]{Department of Physics, University of Oxford, Denys Wilkinson Building, Keble Road, Oxford OX1 3RH, United Kingdom}
\affiliation[e]{{Donostia International Physics Center, Manuel Lardizabal Ibilbidea, 4, 20018 Donostia, Gipuzkoa, Spain}}
\affiliation[f]{{IKERBASQUE, Basque Foundation for Science, 48013, Bilbao, Spain}}

\emailAdd{andrea.rubiola97@gmail.com}

\abstract{We present constraints on the amplitude of matter fluctuations from the clustering of galaxies and their cross-correlation with the gravitational lensing convergence of the cosmic microwave background (CMB), focusing on low redshifts ($z\lesssim0.3$), where potential deviations from a perfect cosmological constant dominating the growth of structure could be more prominent. Specifically, we make use of data from the 2MASS photometric survey (\tmpz) and the \wisc galaxy survey, in combination with CMB lensing data from \planck. Using a hybrid effective field theory (HEFT) approach to model galaxy bias we obtain constraints on the combination $S_8=\sigma_8\sqrt{\Omega_m/0.3}$, where $\sigma_8$ is the amplitude of matter fluctuations, and $\Omega_m$ is the non-relativistic matter fraction. Using a prior on $\Omega_m$ based on the baryon acoustic oscillation measurements of DESI, we find $S_8=0.79\pm0.06$, in reasonable agreement with CMB constraints. We also find that, in the absence of this prior, the data favours a value of $\Omega_m=0.245\pm0.024$,  that is 2.8$\sigma$ lower than \planck.
This result is driven by the broadband shape of the galaxy auto-correlation, and may be affected by theoretical uncertainties in the HEFT power spectrum templates. We further reconstruct the low-redshift growth history, finding it to be compatible with the \planck predictions, as well as existing constraints from lensing tomography. Finally, we study our constraints on the HEFT bias parameters of the galaxy samples studied, finding them to be in reasonable agreement with coevolution predictions.}
\begin{document}
\maketitle

\section{Introduction}\label{sec:intro}
  The growth of fluctuations in the matter density as a function of time is one of the most important cosmological observables, sensitive to the properties of the components that dominate the Universe's expansion, as well as the laws of gravity on cosmological scales \cite{hep-ph/0307284,1003.4231,1902.10503}. Various probes of structure growth exist, such as redshift-space distortions in the three-dimensional clustering of galaxies \cite{1987MNRAS.227....1K}, the amplitude of velocity correlations in peculiar velocity surveys \cite{1708.08236,2105.05185}, correlations in the shapes and positions of galaxies caused by weak gravitational lensing \cite{1010.3829}, the lensing of the cosmic microwave background (CMB) \cite{astro-ph/0601594}, and the non-linear clustering of galaxies \cite{1909.05277,1909.11006}. In this context, the combination of the projected clustering of galaxies at different redshifts, and their cross-correlation with CMB lensing, is a particularly interesting observable. While gravitational lensing is a cumulative effect along the line of sight, the abundance of galaxies is, for the most part, a local tracer of structure. Thus, the correlation between CMB lensing maps and galaxies is sensitive exclusively to the amplitude of matter fluctuations at the redshifts sampled by these galaxies. Combining correlations with galaxies at different redshifts, and using the auto-correlation of these galaxies to constrain their specific clustering properties, we can thus constrain the amplitude of matter fluctuations as a function of time, by ``slicing up'' the CMB lensing map into its contributions at different redshifts. 
  
  This approach, commonly labelled ``CMB lensing tomography'' \cite{2411.08152}, has been used to map out the growth history over a broad range of redshifts. The combination with quasars and other high-redshift samples, has allowed us to reconstruct this history at early times ($z\gtrsim2$) \cite{2103.15862,2402.05761,2409.02109,2506.22416}, and the combination of lensing tomography with the CMB lensing power spectrum is able to push these measurements to even higher redshifts \cite{2409.02109,2507.08798}. The combination with high-density optical and infrared samples, is in turn able to place percent-level constraints at intermediate redshifts ($0.5\lesssim z\lesssim 1.5$) \cite{2105.03421,2105.12108,2309.05659,2407.07949}. Such constraints may be further improved, in both precision and robustness, by the inclusion of higher-order statistics, such as the projected bispectrum \cite{2507.07968}. At low redshifts, below $z\simeq 0.3$, lensing tomography measurements are more scarce, however, due in part to the relatively low lensing signal, and to the significant complication of modelling the clustering of galaxies on the small, non-linear scales, that inevitably project onto large angular scales in this regime \cite{1801.03736,1805.11525,2503.24385}. Targetting these redshifts is nevertheless important, as they may allow us to shed light on the potential tension in the amplitude of matter fluctuations measured by some cosmic shear data sets at late times when compared with the CMB \cite{2007.15633,2105.12108,2105.13543,2105.13544,2304.00701} (although see \cite{2503.19441}), and to search for potential signatures from non-standard dark energy models, as hinted to by the latest BAO and supernova measurements \cite{2503.14738,2504.06118}.

  In this work, we aim to fill this low-redshift gap, by combining high-density galaxy samples at low redshifts with a sophisticated galaxy bias model able to recover robust information from relatively small scales $k\lesssim0.7\,\ihMpc$. Specifically, we will make use of the 2-Micron All-Sky Survey photometric redshift sample (2MPZ, \cite{1311.5246}), the \wisc survey \cite{1607.01182}, concentrated at redshifts $z\lesssim0.4$, and CMB lensing data from the \planck satellite. Recently, \cite{2503.24385} carried out a similar analysis, using low-redshift photometric data from the DESI Legacy Survey, in combination with CMB lensing and galaxy magnification. The results presented here will complement these constraints, sampling this redshift range at higher resolution and pushing towards lower $z$. We will also test the robustness of current constraints to the sample of galaxies used, as well as other analysis choices. Previous constraints using similar data were presented in \cite{1711.04583,1801.03736,1805.11525}, and our analysis will improve upon them in terms of robustness in the modelling of galaxy bias, as well as photometric redshift uncertainties.
  
  This paper is structured as follows. Section \ref{sec:data} presents the galaxy samples and CMB lensing data used. Our theoretical model, as well as the methods used to analysed the data, and to infer cosmological constraints, are described in \ref{sec:method}. The results of our analysis, in terms of both $\Lambda$CDM constraints, growth reconstruction, and galaxy bias properties, are presented in Section \ref{sec:results}. We then present our concluding remarks in Section \ref{sec:conc}.

\section{Data}\label{sec:data}
  \subsection{2MPZ and WIxSC}\label{ssec:data.gals}
    \begin{figure}
        \centering
        \includegraphics[width=0.7\textwidth]{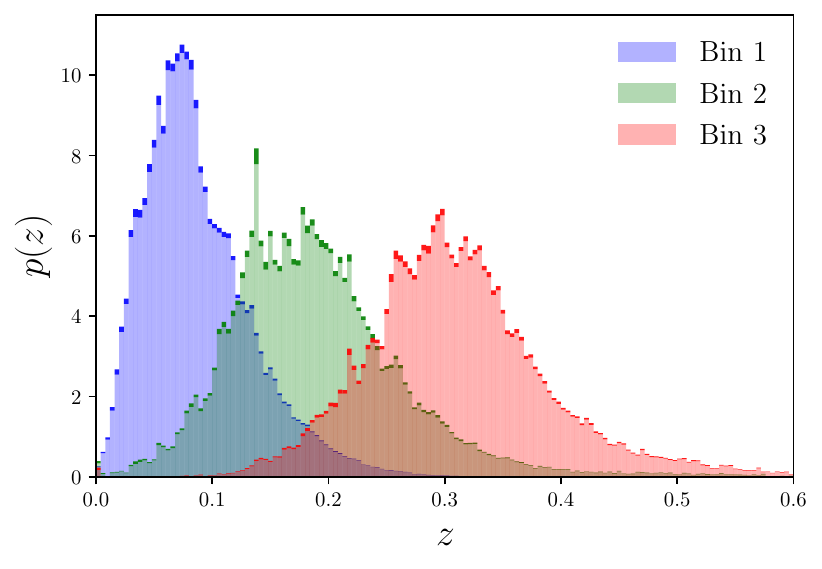}
        \includegraphics[width=0.7\textwidth]{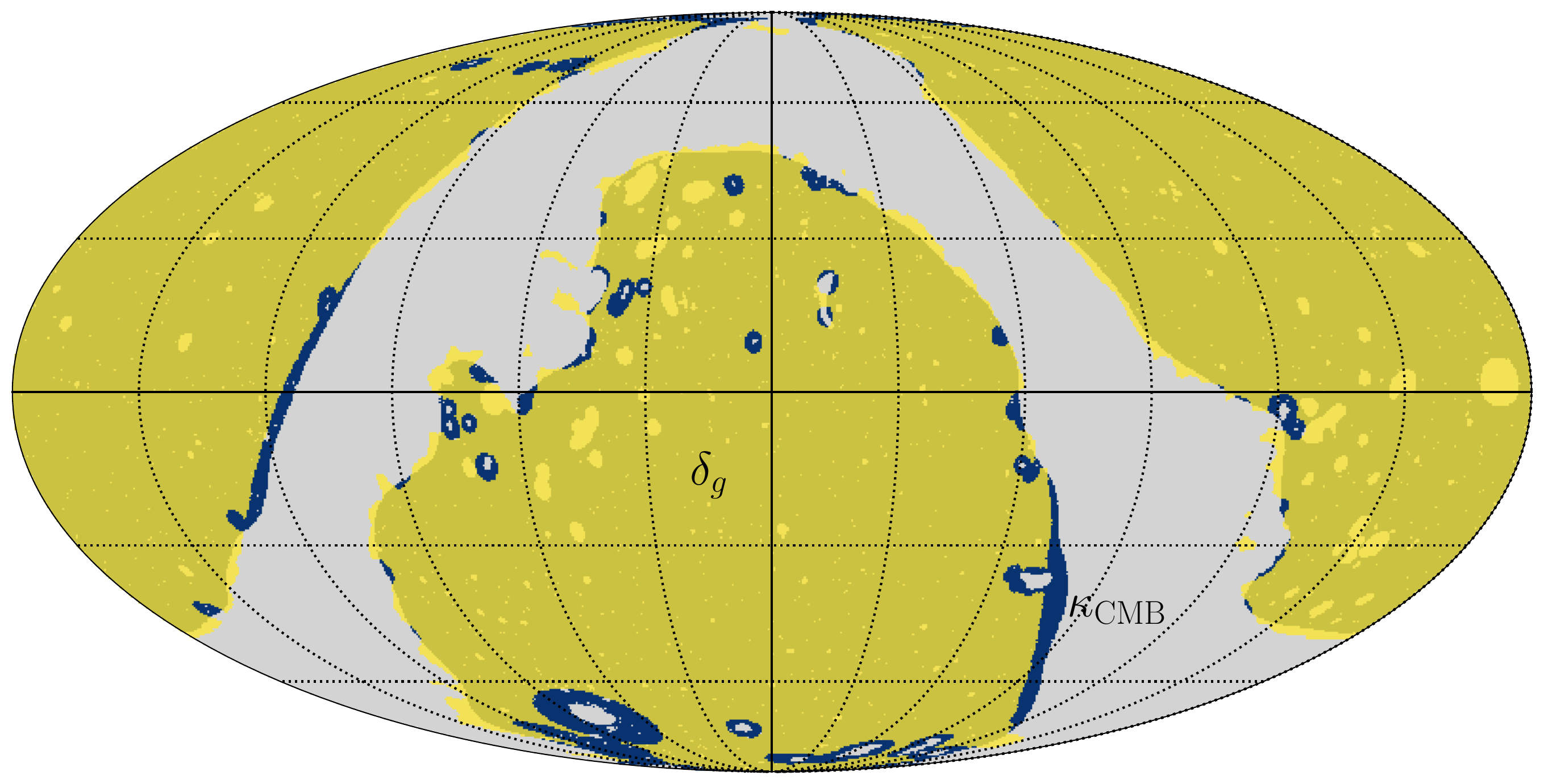}
        \caption{{\sl Top:} redshift distributions of the three galaxy samples considered in this analysis. The bootstrap uncertainties are shown in a darker shade. {\sl Bottom:} sky masks for the galaxy tracers (yellow) and the CMB lensing convergence map (dark blue).}
        \label{fig:dndz}
    \end{figure}
    The 2MASS photometric redshift survey (\tmpz \cite{1311.5246}) combines optical and infrared photometry from the 2-Micron All-Sky Survey (2MASS \cite{astro-ph/0004318}), the Wide-field Infrared Survey Explorer (WISE \cite{1008.0031}), and the SuperCOSMOS survey \cite{astro-ph/0108290}. The resulting magnitude-limited photometric sample (with magnitudes $K_s\leq13.9$) spans the full celestial sphere with highly accurate photometric redshifts ($\sigma_z\simeq0.015$). We use virtually the full 2MPZ sample, selecting galaxies with photometric redshifts $z_p<0.5$, as our lowest-redshift tomographic sample. The sample comprises a total of 718{,}020 galaxies within our analysis mask (see below), with a median redshift $z_m=0.074$ and 90\% of the sample lies at redshifts $z<0.15$. The redshift distribution of the sample was estimated using the direct calibration (DIR) approach of \cite{0801.3822}, using the large spectroscopic sample used to train the 2MPZ photometric redshifts.

    The \wisc survey \cite{1607.01182} was constructed in a manner similar to \tmpz, dropping the 2MASS data in order to select a higher-redshift sample.
    The resulting catalogue, excluding those galaxies present in 2MPZ, contains approximately 20 million sources at redshifts $z\lesssim0.4$, with moderately precise photometric redshifts ($\sigma_z\simeq0.033$). We construct two tomographic samples from \wisc, selecting galaxies with photometric redshifts in the ranges $0.1\leq z_p<0.25$ and $z_p\geq0.25$. These two higher-redshift bins contain 11{,}288{,}037 and 5{,}213{,}088, and lie at median redshifts $z_m=0.18$ and $z_m=0.302$, respectively, although they contain sources up to $z\sim0.5$-$0.6$. The redshift distributions of these samples were calibrated via DIR, using a spectroscopic sample constructed from a variety of existing data sets, as described in \cite{2307.14881}. The uncertainties on these redshift distributions were estimated via bootstrap resampling, and propagated into our final error budget as described in Section \ref{ssec:method.pcl}. Note that we ignore correlated uncertainties between the redshift distributions of the different redshift bins. Fig. \ref{fig:dndz} shows the redshift distributions of the three tomographic bins used here, together with their statistical uncertainties. In what follows we will refer to these samples as ``Bin 1'', 2 and 3, in order of increasing redshift. We construct maps of the galaxy overdensity for each of these samples by first binning each catalogue into sky maps using the \hpx pixelisation scheme with resolution parameter $N_{\rm side}=1024$, corresponding to a pixel size $\delta\theta\simeq3.4'$. This number counts map is first transformed into an overdensity map by subtracting from it the mean number of sources per pixel within our analysis mask, and then dividing the result by it. As discussed in \cite{1909.09102}, the resulting overdensity map is still subject to contamination from stars as well as systematic fluctuations in the different SuperCOSMOS plates. This contamination is limited to multipoles $\ell<10$, which we exclude from our analysis. Nevertheless, we mitigate the residual contamination by subtracting a third-order polynomial function of the local density of stars as mapped by the WISE survey, with the polynomial coefficients estimated by tabulating the observed galaxy overdensity against the star density measured by the WISE survey.

    The analysis mask we use here is described in \cite{1909.09102}, and was constructed by removing the areas most contaminated by Galactic dust and stars, as well as the regions close to the Large and Small Magellanic Clouds. This mask is shown in Fig. \ref{fig:dndz}. The resulting usable sky fraction is $f_{\rm sky}=0.68$, leading to number densities of $\bar{n}=\{25.6,\,402.4,\,185.8\}\,{\rm deg}^{-2}$ for our three redshift bins.
  
  \subsection{Planck CMB lensing}\label{sssec:data.Planck}
    The CMB weak lensing convergence field ($\kappa$), generated by matter overdensities between the last scattering surface and us, receives the greatest contribution from structures located in the $0.5 \lesssim z \lesssim 3$ redshift interval \citep{astro-ph/9810257,astro-ph/0601594}. In this work, we use the convergence map made available as part of the \planck 2018 data release \cite{1807.06210}, covering a sky fraction $\fsky = 0.671$ and thus overlapping spatially with the galaxy samples considered here.

    More precisely,  we will exploit  the ``Minimum Variance'' (MV) lensing convergence harmonic coefficients, which we transform into a \hpx map at $N_{\rm side}=1024$ resolution after filtering out the harmonic coefficients with $\ell> 3 N_{\rm side} - 1 = 3071$. We use the binary sky mask made available with this map. Since the lensing reconstruction noise power spectrum rises sharply with $\ell$ on small scales, we apodise this mask with a $0.2^\circ$ ``C1'' kernel \cite{1809.09603} in order to minimise the leakage from noise-dominated small-scale modes. The final usable sky fraction is $\fsky \simeq 0.66$. At the angular power spectrum level, following \cite{1502.01591,1807.06210}, we remove the smallest multipoles $\ell < 8$ , which are too sensitive to the mean-field subtraction in the lensing reconstruction process. Similarly, we use up to $\ell_{\rm max} = 2000$, corresponding to the \textit{aggressive} scale ranges of \cite{1807.06210}.

    Our analysis will not include the CMB lensing auto-correlation signal: we thus do not need to include a rigorous modelling of the various noise bias terms that enter the CMB lensing likelihood. Yet, the CMB lensing noise enters the covariance matrix for any power spectra involving the CMB lensing map. Its contribution will be taken into account through our estimate of the power spectrum covariance, as described in Section \ref{ssec:method.pcl}.

\section{Methodology}\label{sec:method}
  \subsection{Projected statistics: galaxy clustering and CMB lensing}\label{ssec:method.cls_th}
    We will explore two projected tracers of the large-scale structure, the angular galaxy overdensity $\delta_g$ (in three different redshift bins) and the CMB lensing convergence $\kappa$. Both of these tracers can be written in general as 
    \begin{equation}
      a(\nv)=\int d\chi\,q_a(\chi)\,A(\chi\nv,z(\chi)),
    \end{equation}
    where $\nv$ is a unit vector along the line of sight, $\chi$ is the radial comoving distance, and $z$ is the corresponding redshift in the background. Here, $A$ is the three-dimensional field being traced by the projected quantity $a$, and $q_a(\chi)$ is a radial kernel. The radial kernel for $\delta_g$ is simply proportional to the redshift distribution of the galaxy sample, and the CMB lensing kernel covers all redshifts from $z=0$ to the epoch of recombination, peaking at $z\sim2$:
    \begin{equation}
      q_g(\chi)=H(z)\,p(z),\hspace{12pt}
      q_\kappa(\chi)=\frac{3}{2}\Omega_m\,H_0^2\,(1+z)\,\chi(z)\left(1-\frac{\chi(z)}{\chi_{\rm LSS}}\right)\,\Theta(z<z_{\rm LSS}).
    \end{equation}
    Here $H(z)$ is the expansion rate\footnote{Note that we use natural units, with $c=1$, throughout.}, $H_0\equiv H(0)$, $\chi_{\rm LSS}$ is the distance to the surface of last scattering, at redshift $z_{\rm LSS}=1100$, and $\Theta$ is the Heavyside function. We assume a flat $\Lambda$CDM model throughout.
    
    The three-dimensional fields probed by both tracers are the 3D galaxy overdensity $\Delta_g$, and the matter overdensity $\Delta_m$ in the case of $\kappa$. The angular power spectrum of any two such projected fields, $a$ and $b$, can be related to the 3D power spectrum of their associated 3D quantities $P_{AB}(k)$ via
    \begin{equation}
      C_\ell^{ab}=\int \frac{d\chi}{\chi^2}\,q_a(\chi)\,q_b(\chi)\,P_{AB}\left(k=\frac{\ell+1/2}{\chi}, z(\chi)\right)
    \end{equation}
    This relation makes use of the Limber approximation \cite{Limber_1953}. We have tested that adopting this approximation does not significantly change our results, modifying our constraints on e.g. $S_8$ by less than $\sim2\%$ (significantly less than our statistical uncertainties). To quantify this, we used the simplified bias model introduced in Appendix \ref{app:om} to derive constraints on $\Omega_m$ and $\sigma_8$ (the latter from the ratio between the scale-independent amplitude parameters). This simplified model allows for a quick evaluation of the likelihood, since the angular power spectrum templates (which is made significantly slower when discarding the Limber approximation) must only be recalculated when varying $\Omega_m$.

    We also neglect the impact of lensing magnification on the measured spectra, owing to the relatively low redshifts of the samples explored here. The effect is significantly smaller than our statistical uncertainties even in a conservative scenario, assuming a large value for the logarithmic slope in galaxy counts for this sample.

    Our analysis will be based on measurements of the galaxy auto-correlation $C_\ell^{gg}$ and $C_\ell^{g\kappa}$. Hence, the last ingredient needed to make theoretical predictions is a model for the 3D galaxy-galaxy and galaxy-$\kappa$ power spectra.
  
  \subsection{Perturbative bias expansion}\label{ssec:method.heft}
    We model the galaxy power spectra using a hybrid Lagrangian bias expansion, also called Hybrid Effective Field Theory (HEFT) \citep{0807.1733,1910.07097,2101.12187,2101.11014,2103.09820}. In this context the galaxy overdensity field can be expanded (in Lagrangian coordinates) as a function of operators acting on the underlying Lagrangian matter field. Stopping the expansion at second order, the only operators that do not break any symmetries are $\delta, \delta^2, s^2$ and $\nabla^2\delta$, where $\delta$ is the linear matter density and $s^2$ is the traceless contracted tidal tensor \citep{1910.07097}. Moreover, observed quantities are in the final, Eulerian coordinates, which requires advecting the Lagrangian galaxy density obtained to its late-time coordinates. The final galaxy overdensity field in Eulerian coordinates $\boldsymbol{x}$ therefore reads
    \begin{equation}
      1 + \Delta_g (\boldsymbol{x}) = \int \mathrm{d}^3 q\,[1 + b_1 \delta(\boldsymbol{q}) + \dfrac{b_2}{2}\left[ \delta^2(\boldsymbol{q})  - \langle \delta^2 \rangle \right]+ \dfrac{b_{s}}{2} \left[s^2(\boldsymbol{q})- \langle s^2 \rangle \right]+ b_{\nabla^2} \nabla^2\delta(\boldsymbol{q})] \delta_{\rm D}\left(\boldsymbol{x} - \boldsymbol{q} - \boldsymbol{\Psi}(\boldsymbol{q}) \right),
      \label{eq::advection}
    \end{equation}
    where $b_1, b_2, b_{s^2},$ and $b_{\nabla^2}$ are free parameters (the galaxy bias parameters),  $\boldsymbol{q}$ are the Lagrangian coordinates, and $\boldsymbol{\Psi}$ is the displacement vector connecting the Lagrangian to the Eulerian coordinates.   
While in a fully perturbative approach $\boldsymbol{\Psi}$ can be obtained from an expansion of the density field, in the HEFT framework this is instead calculated through a fully non-linear $N$-body solver. The corresponding galaxy auto-power spectrum and galaxy-matter cross-power spectrum are
    \begin{equation}
    \begin{split}
        P_{gg}(k) = \sum_{i,j} b_i b_j P_{ij}(k),\hspace{12pt} P_{gm}(k) = \sum_{i} b_i P_{1i}(k),
    \end{split}
    \end{equation}
    where the indices $i, j$ run on the different operators entering the expansion, and $P_{ij}(k)$ is the cross-power spectrum of the $i$-th and $j$-th advected operators. The specific implementation of HEFT we use is the one presented in \citep{2101.12187}, where the cross spectra $P_{ij}$ are measured in the BACCO $N$-body simulations \citep{2004.06245,2104.14568} and their evaluation as a function of cosmology and redshift is accelerated through the use of the neural-network-based emulator \bemu \cite{Zennaro_coevolution}.

    HEFT has been shown to provide unbiased predictions for the 2-point clustering of a wide range of galaxy types up to relatively small scales, $k\lesssim 0.7\,h{\rm Mpc}^{-1}$ \cite{2110.05408,2307.03226,2312.00679}. Furthermore, it has been found that the values of the higher-order bias coefficients, $b_2$, $b_s$, $b_{\nabla^2}$ tend to be correlated with the linear bias $b_1$, around the so-called ``coevolution relations'' \citep{2110.05408}, albeit with a relatively broad scatter.
  
  \subsection{Power spectra and covariances}\label{ssec:method.pcl}
    We estimate all power spectra and their associated covariance matrix using the pseudo-$C_\ell$ approach \cite{Hivon_2002,1809.09603}. In this framework, the observed version of any projected field $a^v$ is a product, in real space, of the true underlying field $a$ and its mask $v$:
    \begin{equation}
      a^v(\nv)=v(\nv)\,a(\nv).
    \end{equation}
    This relationship transforms into a convolution in harmonic space, and a similar result holds then for the ensemble average of the power spectra of two masked field,  $a^v$ and $b^w$, and their masks, $v$ and $w$:
    \begin{equation}
      \langle \tilde{C}^{ab}_\ell\rangle=\sum_{\ell '}M^{vw}_{\ell\ell'}\,C_\ell^{ab}+\tilde{N}^{ab}_\ell.
    \end{equation}
    Here, $C_\ell^{ab}$ is the underlying angular power spectrum of the fields, and $\tilde{C}^{ab}_\ell$ is the so-called ``pseudo-$C_\ell$'', or ``coupled $C_\ell$'':
    \begin{equation}
      \tilde{C}^{ab}_\ell\equiv\frac{1}{2\ell+1}\sum_{m=-\ell}^\ell {\rm Re}(a^v_{\ell m}(b^w_{\ell m})^*).
    \end{equation}
    $\tilde{N}_\ell^{ab}$ is the noise pseudo-spectrum (i.e. the spectrum of the noise contributions to $a$ and $b$, typically mostly relevant for auto-correlations). Finally, $M^{vw}_{\ell\ell'}$ is the mode-coupling matrix, describing the statistical correlation between different angular multipoles $\ell$ caused by the presence of the mask. Crucially, $M^{vw}_{\ell\ell'}$ depends solely on the pseudo-$C_\ell$ of the two masks, and can be estimated analytically in a computationally efficient manner. Finally, the quantity we use in our inference is the ``decoupled'' bandpowers, obtained by binning the pseudo-$C_\ell$ and inverting the resulting binned mode-coupling matrix (see \cite{1809.09603} for further details). We use the mode-coupling matrix to estimate the ``bandpower window functions'', which connect the theoretical predictions of the angular power spectra described in the previous section with the measured bandpowers.
    
    We use an analytical estimate of the covariance matrix consisting of four contributions:
    \begin{eqnarray}
      {\rm Cov}={\rm Cov}_{\rm G}+{\rm Cov}_{\rm cNG}+{\rm Cov}_{\rm SN}+{\rm Cov}_{p(z)}.
    \end{eqnarray}
    These contributions are:
    \begin{itemize}
      \item ${\rm Cov}_{\rm G}$: the disconnected Gaussian covariance, which dominates the total uncertainty. We estimate this using the improved Narrow Kernel Approximation (iNKA) as described in \cite{1906.11765,Nicola_2020}, which has been shown to accurately recover the correlations between different angular scales induced by survey geometry.
      \item ${\rm Cov}_{\rm cNG}$: the connected non-Gaussian trispectrum contribution. We estimate this component using the halo model as described in \cite{1601.05779}. For this, we first fit the angular auto-correlations of each redshift bin using a Halo Occupation Distribution (HOD) model \cite{astro-ph/0005010} in the range of scales $k<1\,h\,{\rm Mpc}^{-1}$. We use the HOD parametrisation of \cite{astro-ph/0408564}, in which the mean number of central and satellite galaxies is given by
      \begin{align}\nonumber
        \bar{N}_c(M)=\frac{1}{2}\left[1+{\rm erf}\left(\frac{\log(M/M_{\rm min})}{\sigma_{{\rm ln}M}}\right)\right], \hspace{12pt}
        \bar{N}_s(M)=\Theta(M-M_0)\,\left(\frac{M-M_0}{M_1}\right)^\alpha.
      \end{align}
      We fix $\alpha=1$, $M_0=M_{\rm min}$, $\sigma_{{\rm ln}M}=0.4$, and treat $M_{\rm min}$ and $M_1$ as free parameters. The cNG contribution is subdominant on the scales explored in this analysis.
      \item ${\rm Cov}_{\rm SN}$: the contribution to the covariance from uncertainties in the effective shot noise of the sample. Although we correct all galaxy auto-spectra for the shot-noise contribution associated with the discreteness of the sample analytically, other stochastic contributions to the galaxy overdensity may lead to additional terms in the auto-spectrum that are effectively white (i.e. uncorrelated on the scales probed). We thus marginalise over an additional amplitude parameter $A_{\rm SN}^i$ of the shot noise contribution to the auto-spectrum of the $i$-th bin to account for these effects. Since the theory model is linear in this parameter, we can marginalise over it analytically by adding a contribution to the power spectrum covariance of the form:
      \begin{equation}
        {\rm Cov}_{\rm SN}(C_\ell^{gg},C^{gg}_{\ell'})=\sigma_{\rm SN}^2\,N_\ell^{gg}\,N_{\ell'}^{gg},
      \end{equation}
      where $N_\ell^{gg}$ is the shot noise contribution to the power spectrum, estimated as described in e.g. \cite{1912.08209}, and $\sigma_{\rm SN}$ is the standard deviation of the Gaussian prior assumed on $A_{\rm SN}$. We use $\sigma_{\rm SN}=0.1$, thus allowing for $\sim10\%$ deviations from a pure Poisson contribution. We tested that our results vary only mildly, with parameter shifts below $0.5\sigma$, if increasing this prior significantly to $\sigma_{\rm SN}=1$. We include these terms in the diagonal covariance blocks corresponding to each of the galaxy auto-correlations (i.e. effectively marginalising over a shot noise amplitude parameter for each redshift bin).
      \item ${\rm Cov}_{p(z)}$: the contribution to the covariance from uncertainties in the redshift distributions of the three galaxy samples. We propagate the uncertainties on the redshift distributions, estimated as described in Section \ref{ssec:data.gals}, following the approach outlined in \cite{2007.14989} and validated in \cite{2301.11978}. The structure of the resulting covariance contribution is similar to that of ${\rm Cov}_{\rm SN}$:
      \begin{equation}
        {\rm Cov}_{p(z)}={\sf T}^T{\sf P}{\sf T},
      \end{equation}
      where $({\sf T})_{\alpha j}\equiv \partial t_\alpha/\partial [p(z)]_j$ is the derivative of the $\alpha$-th element of our model prediction for the full data vector with respect to the $j$-th element in the list of uncertain redshift distribution elements, and $({\sf P})_{ij}$ is the Gaussian prior covariance of the redshift distributions. We construct ${\sf P}$ from the bootstrap uncertainties calculated in Section \ref{ssec:data.gals} assuming uncorrelated errors.
    \end{itemize}
  
  \subsection{Likelihood}\label{ssec:method.like}
    \begin{table}[t]
      \centering
      \large
      \renewcommand{\arraystretch}{1.1} 
      \begin{tabular}{|c c|}
        \hline
        \textbf{Parameter} & \textbf{Prior} \\
        \hline
        $\sigma_8$ & $U(0.4,1.2)$ \\
        $\Omega_m$  & $U(0.1,0.5)$ or $\mathcal{N}(0.2975,0.0086)$ \\
        $b_{1,i},\,b_{2,i}\,b_{s,i}$ & $\mathcal{N}(0,5)$ \\
        $b_{\nabla^2,i}$ & $\mathcal{N}(0,10\,{\rm Mpc}^2)$ \\
        \hline
      \end{tabular}
      \caption{Priors on the full set of parameters employed in our analysis. $U(a,b)$ stands for an uniform distribution between $a$ and $b$, and $\mathcal{N}(a, b)$ for a normal distribution with mean $a$ and variance $b$. In the case of $\Omega_m$, we explore both a non-informative flat prior, and an informative prior based on the BAO constraints from DESI \cite{2503.14738}.}
      \label{tab:priors_cosmo_bias}
    \end{table}
    \subsubsection{Likelihood, priors, and scale cuts}\label{sssec:method.like.like}
      Having marginalised over the $p(z)$ uncertainties and deviations from Poisson shot noise, our model depends explicitly on the HEFT galaxy bias parameters of each redshift bin and on cosmological parameters. To derive constraints on the free parameters of the model we will assume a Gaussian likelihood, appropriate for power spectrum measurements on the scales explored here (particularly given the wide sky area covered by the data sets used) \cite{0801.0554}. In this case:
      \begin{eqnarray}
        -2\log p({\bf d}|\vec{\theta})=\left({\bf d}-{\bf t}(\vec{\theta})\right)^T\,{\rm Cov}^{-1}\,\left({\bf d}-{\bf t}(\vec{\theta})\right)+K,
      \end{eqnarray}
      where ${\bf d}$ is the data vector (comprised in our case of galaxy auto-spectra and galaxy-lensing cross-spectra for different redshift bins),  ${\bf t}$ is the theory prediction, dependent on the model parameters $\vec{\theta}$, and constructed using the model described in Sections \ref{ssec:method.cls_th} and \ref{ssec:method.heft}, ${\rm Cov}$ is the covariance matrix of ${\bf d}$, estimated as described in Section \ref{ssec:method.pcl}, and $K$ is an irrelevant normalisation constant.

      To construct the posterior distribution we assume the following priors on the model parameters (summarised in Table \ref{tab:priors_cosmo_bias}): we assume Gaussian priors for the linear, quadratic, and tidal bias parameters, centred at 0 with a standard deviation of 5. In turn, we use a Gaussian prior for the non-local bias term $b_{\nabla^2}$ with zero mean and a standard deviation of $10\,{\rm Mpc}^2$. These Gaussian priors have the advantage of allowing us to use the approximate analytic marginalisation method (AAM) briefly described in Section~\ref{sssec:method.like.prof} while being broad enough to avoid driving our constraints. We consider variations in two cosmological parameters, $\sigma_8$ and $\Omega_m$. We use a uniform flat prior for $\sigma_8$, and we will consider two cases for $\Omega_m$:
      \begin{itemize}
        \item Free $\Omega_m$: we impose a large flat prior in the range $0.1\leq\Omega_m\leq0.5$. This will allow us to explore the values of $\Omega_m$ preferred by our data. As we will see, these constraints will be driven by the broadband shape of the galaxy auto-spectrum.
        \item $\Omega_m$ constrained by BAO: we impose a Gaussian prior on $\Omega_m$ using the BAO constraints found by the DESI collaboration \cite{2503.14738}: $\Omega_m=0.2975\pm0.0086$. This allows us to break the partial degeneracy existing between $\sigma_8$ and $\Omega_m$ when analysing projected clustering and lensing data, and related to the uncertainty in the comoving distance to the galaxy samples under study \cite{2022JCAP...02..007W}. It will also allow us to avoid potential error when extrapolating the HEFT power spectrum templates to very low values of $\Omega_m$, outside the \bemu training range (see Section \ref{sssec:method.like.extrap}). The resulting constraints on $\sigma_8$ will still be driven by low-redshift information, while employing a more robust observable to constrain $\Omega_m$ than the broad-band shape of the power spectrum.
      \end{itemize}
      We will also report constraints on the derived parameter $S_8\equiv\sigma_8\sqrt{\Omega_m/0.3}$, which is substantially less degenerate with $\Omega_m$. We fix all other parameters of the $\Lambda$CDM model. We consider massless neutrinos ($\sum m_\nu=0\,{\rm eV}$), in order to significantly speed up the likelihood evaluation. We checked that our cosmological constraints vary by less than 1\% (or $0.2\sigma$ in $S_8$) if setting the sum of neutrino masses to the lower bound from neutrino oscillation experiments $\sum m_\nu=0.06\,{\rm eV}$ instead. We fixed the physical baryon density parameter and the scalar spectral index to the values inferred by \planck $\Omega_bh^2=0.02236$, $n_s=0.9649$, and we fix $\Omega_mh^3= 0.09633$ as a proxy for the angular acoustic scale. The latter constrain effectively controls the value of the Hubble constant $h$ for a given choice of $\Omega_m$.

      When analysing our data using the HEFT bias model we consider only measurements corresponding to physical scales $k<k_{\rm max}^{\rm HEFT}=0.42\,{\rm Mpc}^{-1}$ (corresponding to $\sim 0.6\,h{\rm Mpc}^{-1}$ for $h\simeq0.7$), where HEFT is known to remain accurate \citep{2101.11014,2101.12187,2103.09820,2207.06437,2307.03226,2405.02252}. We impose these scale cuts by translating this $k_{\rm max}^{\rm HEFT}$ into a maximum angular multipole $\ell_{\rm max}\equiv\chi(\bar{z})\,k_{\rm max}^{\rm HEFT}$, where $\chi(\bar{z})$ is the comoving distance to the mean redshift of a given galaxy sample in the best-fit \planck cosmology. We also remove the largest angular scales, with $\ell<\ell_{\rm min}=20$ to avoid biases in the galaxy auto-correlation from large-scale systematics, which are particularly relevant for the \wisc sample \cite{1607.01182}. Note that, when cross-correlating galaxies and CMB lensing, the  galaxy scale cuts will dominate, as they are more conservative than those in Section~\ref{sssec:data.Planck}.

    \subsubsection{HEFT emulator extrapolation}\label{sssec:method.like.extrap}
      The flat priors we assume on $\Omega_m$ and $\sigma_8$ are broader than the parameter range over which the \texttt{baccoemu} emulator for the HEFT power spectrum templates has been trained. In order to obtain reliable estimates of the HEFT templates outside this range we extrapolate beyond this range using a controlled Taylor expansion at first order. Specifically, for a given point $\vec{\theta}$ outside of the \texttt{baccoemu} training range we estimate the $P_{ij}$ template as
      \begin{eqnarray}
        P_{ij}(k, \theta) = P_{\rm mm}(k, \vec{\theta}) \left( {\cal R}_{ij}(k,\vec{\theta}_b) +  \nabla_{\vec{\theta}}{\cal R}_{ij}\rvert_{k,\vec{\theta_b}}\cdot(\vec{\theta} - \vec{\theta}_b) \right),
      \end{eqnarray}
      where $\vec{\theta}_b$ is the closest point to $\vec{\theta}$ within the training range, and we have defined the ratio
      \begin{eqnarray}
        {\cal R}_{ij}(k,\vec{\theta})\equiv\frac{P_{ij}(k,\vec{\theta})}{P_{mm}(k,\vec{\theta})}.
      \end{eqnarray}
      In these expressions $P_{mm}(k,\vec{\theta})$ is the matter power spectrum. This procedure ensures that the extrapolation is carried out only on the ratio ${\cal R}_{ij}$, which has a milder dynamic range and parameter dependence than $P_{ij}$ (e.g. the dependence on $\sigma_8$ is largely absorbed by the ratio with $P_{mm}$), as well as a smoother scale dependence. To ensure that $P_{mm}(k,\vec{\theta})$ can be evaluated outside the \bemu training range, we calculate it using the {\tt HALOFIT} parametrisation \cite{1208.2701}, with the linear power spectrum calculated using the fast parametrisation of \cite{astro-ph/9709112} to ensure a fast evaluation of the likelihood. We tested that resorting to a Boltzmann solver instead led to sub-percent variations in the recovered cosmological constraints. Finally, to further improve the quality of the extrapolated templates, we modify the definition of ${\cal R}_{ij}$ for the HEFT templates that are positive-definite (e.g. $P_{\delta^2\delta^2}(k)$), replacing it by its logarithm. I.e. in these cases we use
      \begin{eqnarray}
        P_{ij}(k, \theta) = P_{\rm mm}(k, \vec{\theta}) \exp\left[\left( \tilde{\cal R}_{ij}(k,\vec{\theta}_b) +  \nabla_{\vec{\theta}}\tilde{\cal R}_{ij}\rvert_{k,\vec{\theta_b}}\cdot(\vec{\theta} - \vec{\theta}_b) \right)\right],
      \end{eqnarray}
      with $\tilde{\cal R}_{ij}\equiv \log{\cal R}_{ij}$.
      
      \begin{figure}
        \centering
        \includegraphics[width=0.75\textwidth]{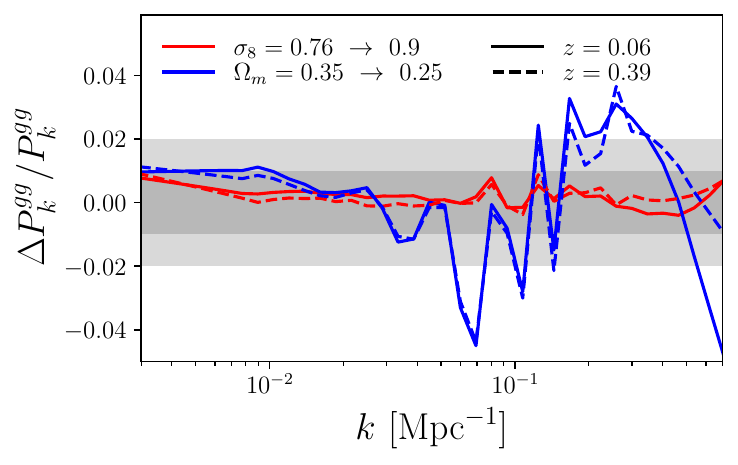}
        \caption{Relative error in the galaxy power spectrum incurred when extrapolating the dependence of the HEFT power spectrum templates on cosmological parameters. The red and blue lines show the result of extrapolating in $\sigma_8$ and $\Omega_m$, respectively. For guidance, the gray bands mark the limits where relative variations would exceed 1\% and 2\%, comparable with the sensitivity of our galaxy auto-correlation measurements}. The start and end points of the extrapolation in each parameter are shown in the legend. Solid and dashed lines show results at $z=0.06$ and $0.39$, respectively.
        \label{fig:extrap_heft}
      \end{figure}

      To quantify the accuracy of this extrapolation scheme we compare the exact value of the galaxy power spectrum at one end of the \textsc{baccoemu} training range against the same power spectrum extrapolated from the opposite end of the range using the method above. In these cases, when constructing the total galaxy power spectrum we use a value of $1+b_1=1.3$ consistent with the large-scale bias of the samples studied here, and set the higher-order bias parameters to the values corresponding to the coevolution relations of \cite{2110.05408} for this value of $b_1$. As an example, Fig. \ref{fig:extrap_heft} shows the relative differences between the true and extrapolated power spectra when extrapolating from low to high values of $\sigma_8$ and from high to low values of $\Omega_m$. We find that the extrapolation in $\sigma_8$ is remarkably stable, achieving sub-percent accuracy on most scales. The extrapolation in $\Omega_m$ is less robust, however, leading to differences of up to $\sim4\%$, particularly around the BAO scale ($k\sim0.1\,{\rm Mpc}^{-1}$). Although this is likely sufficiently accurate given the precision of our measurements, the constraints we derived when imposing a BAO prior on $\Omega_m$ will be more robust against potential inaccuracies in the parameter dependence of the HEFT bias templates outside the \bemu training range.

    \subsubsection{Volume effects and profile likelihood}\label{sssec:method.like.prof}
      It has been shown that models based on perturbative bias expansions may suffer from so-called ``volume'' or ``projection effects'', significant shifts in the marginalised posterior constraints on cosmological parameters resulting from degeneracies with the large number of free bias parameters. These volume effects may lead to incorrect conclusions regarding, for example, the level of tension between large-scale structure and CMB constraints on growth \cite{2301.11895,2404.07312}. Various approaches have been proposed in the literature to minimise these effects, including the use of simulation-based priors on the bias parameters \cite{2402.13310,2507.00118}, the use of well-educated reparametrisations \cite{2509.09562}, or the use of Jeffreys priors \cite{2301.11895}. Here we will employ the approximate analytic marginalisation (AAM) procedure outlined in \cite{2301.11895} \textcolor{red}{\citep{2026JCAP...03..062D}}, based on the profile likelihood. As shown in \cite{2301.11895}, the result of AAM is a marginalised likelihood for the cosmological parameters alone where, by construction, the marginalised posterior distributions are centred at the best-fit model parameters (and hence are free from volume effects), and which is approximately equivalent to the use of a Jeffreys prior for the bias parameters. AAM also allows for a much faster exploration of large parameter spaces, which is relevant in our case when combining all redshift bins. In this case our model depends on 14 model parameters, 12 of which are nuisance bias parameters. We use this method as implemented in the {\tt Cosmotheka} likelihood\footnote{\url{https://github.com/Cosmotheka/Cosmotheka_likelihoods/tree/main/ClLike}}, which interfaces with \texttt{cobaya} \cite{2005.05290} to sample the posterior distribution, and makes use of the Core Cosmology Library \cite{1812.05995} for most theoretical calculations. To ensure converged results, we perform the Gelman-Rubin test using multiple chains, and stop the sampling when $R-1=0.01$.
      
\section{Results}\label{sec:results}
  \subsection{Power spectra and constraints on $\Lambda$CDM}\label{ssec:results.lcdm}
    \begin{figure}[t]
        \centering
        \includegraphics[width=1.\linewidth]{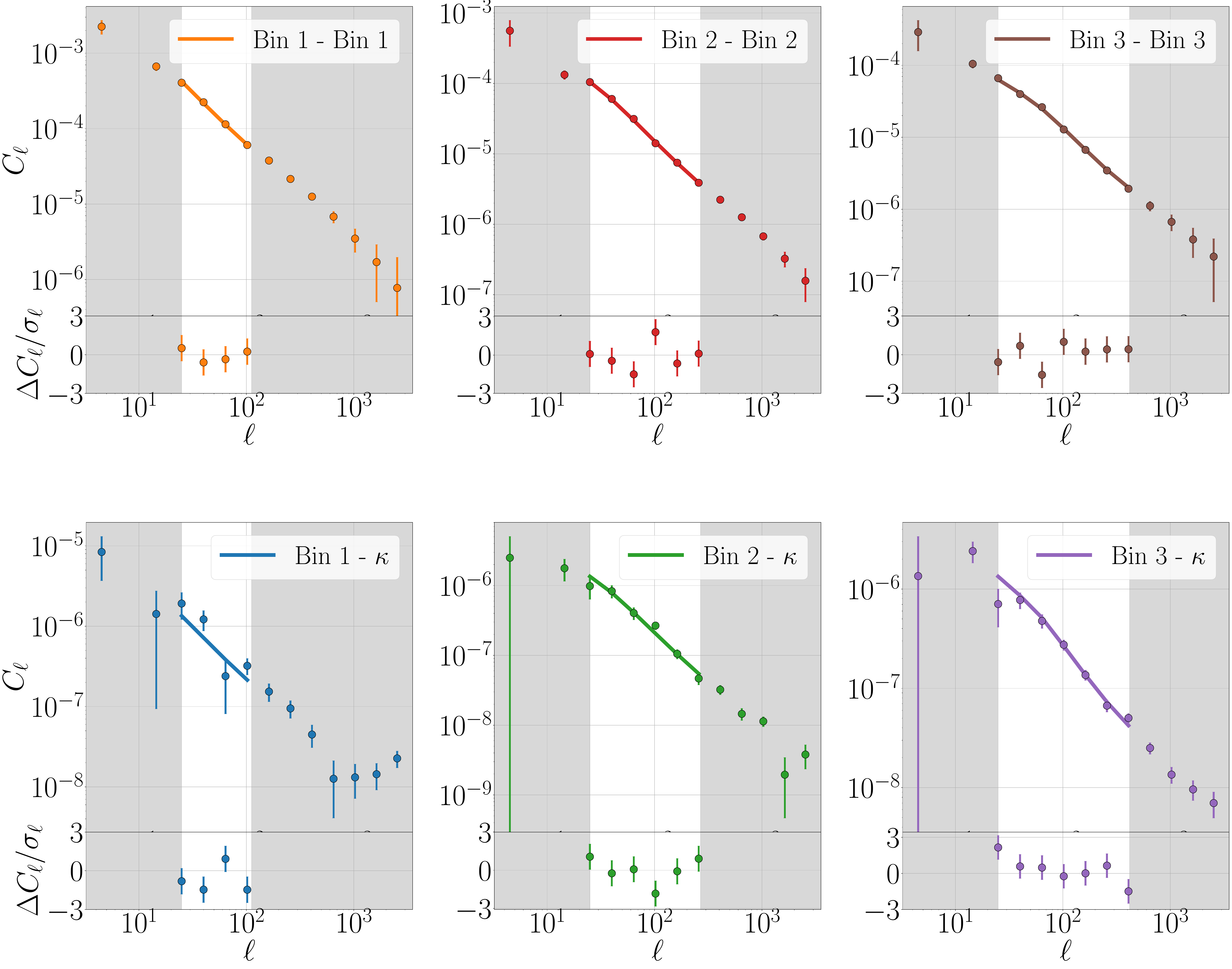}
        \caption[Power spectrum measurements and best-fit model predictions.]{Power spectrum measurements and best-fit theory predictions. Each column displays the galaxy auto-correlation $C^{gg}_\ell$ (top) and the cross-correlation with CMB lensing $C^{g\kappa}_\ell$ (bottom), with the results for Bins 1, 2, and 3 (left, center, right). The sub-panel in each figure shows the difference between the data and the fiducial best-fit model (shown as a solid line in the main panels) as a fraction of the statistical uncertainties. The shaded bands show the angular scales excluded from the analysis.}
        \label{fig:cl_plot_wisc_all}
    \end{figure}
    Fig. \ref{fig:cl_plot_wisc_all} shows the power spectrum measurements used in this analysis, together with the best-fit theory predictions found within our fiducial $\Lambda$CDM analysis (described below). Within the scale cuts used here, we find that the model provides a relatively good fit to the data. The galaxy-lensing cross-correlation is detected at relatively high significance in all redshift bins. Within the scales used for our analysis, $C_\ell^{g\kappa}$ is measured with a signal-to-noise ratio of 6.5, 14, and 19 in bins 1, 2, and 3 respectively. Note that, due to the low mean redshift of Bin 1, only four data points survive our scale cuts in both $C_\ell^{gg}$ and $C_\ell^{g\kappa}$, which will severely limit the amount of information we can extract from this bin.

    \begin{figure} 
      \centering
      \includegraphics[width=0.7\linewidth]{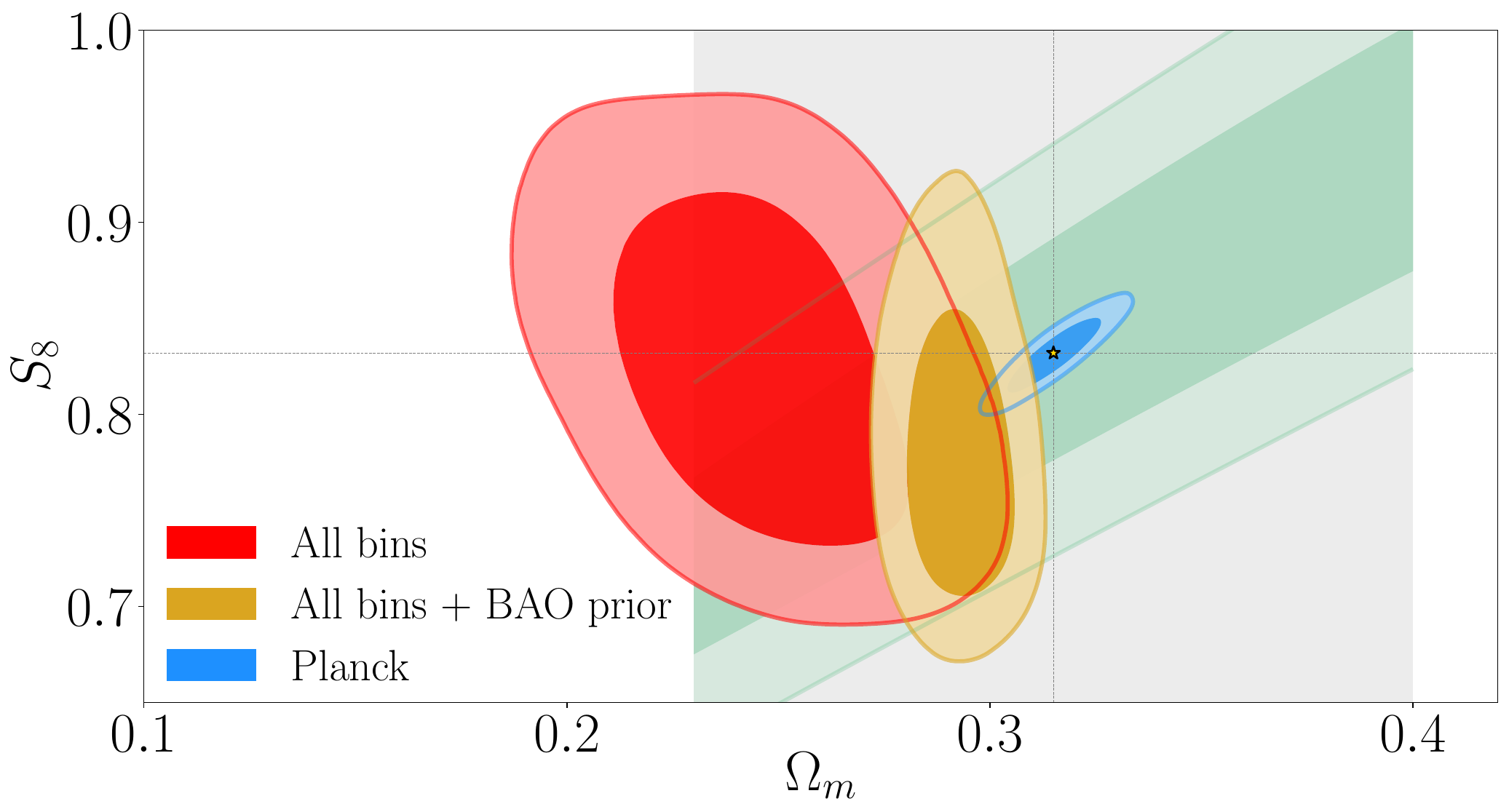}
      \caption[Constraints on $S_8$ and $\Omega_m$]{Two-dimensional 68\% and 95\% constraints in the $(\Omega_m,S_8)$ plane. Results are shown for our fiducial analysis, combining all redshift bins, with $\Omega_m$ as a free parameter (red contours) and imposing a BAO prior from DESI on $\Omega_m$ (yellow), which we compare to \planck (blue). The gray band shows the $\Omega_m$ range over which the \bemu HEFT emulator has been trained. The light-green oblique band shows the range of $S_8$ values corresponding to the emulator's training range in $\sigma_8$ as a function of $\Omega_m$.}\label{fig:baseconfig_and_Planck.pdf}
    \end{figure}

\begin{figure}
    \centering
    \includegraphics[width=1\linewidth]{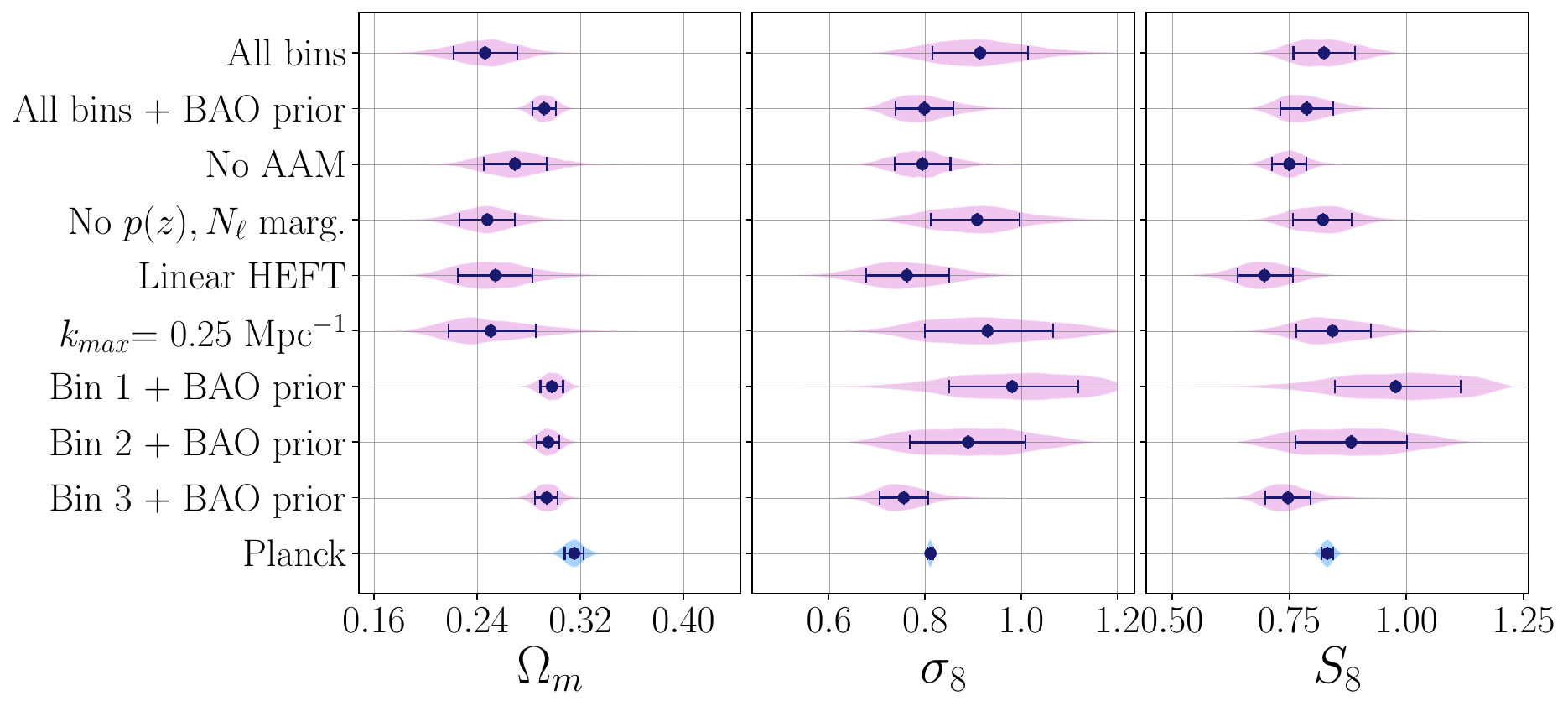}
    \caption[Violin plot for $\Omega_m$,$S_8$ and $\sigma_8$]{Constraints on $\Omega_m$, $S_8$ and $\sigma_8$ found for the different analysis choices explored here, as well as the \planck measurements. The numerical constraints in each case are shown in Table \ref{tab:constraints_bestfit}. The black points and error bars represent the mean and $68\%$ confidence intervals, the same statistics used to estimate the tension levels. The pink shapes represent the marginalised posterior distributions from the MCMC chains.}
    \label{fig:all_violins}
\end{figure}

    Fig. \ref{fig:baseconfig_and_Planck.pdf} shows constraints on $\Omega_m$ and $S_8$ from the combination of $C_\ell^{gg}$ and $C_\ell^{g\kappa}$ measured in our three redshift bins on scales up to $k_{\rm max}=0.42\,{\rm Mpc}^{-1}$. Results are shown for the case in which $\Omega_m$ is left completely free (red contour) and after including the DESI BAO prior (yellow contours). For comparison, the \planck constraints on these parameters are shown in blue. The numerical constraints on $\Omega_m$, $\sigma_8$, and $S_8$ are provided in Table \ref{tab:constraints_bestfit}, which also includes the constraints found for all other alternative analysis choices explored here. These results can be visually compared in Fig. \ref{fig:all_violins}. Our constraints on $S_8$ in either case are:
    \begin{eqnarray}
      S_8=0.82\pm0.07\,\,{\rm (free}\,\,\Omega_m),
      \hspace{12pt}
      S_8=0.79\pm0.06\,{\rm (DESI\,\,BAO\,\,prior)}.
    \end{eqnarray}
    
  \begin{table}
  \small
\centering
\begin{tabular}{|l|c|c|c|c|}
\hline
\textbf{Configuration} ($N_{\rm dof,eff}$) & $\Omega_m$ (b.f.) & $S_8$ (b.f.) & $\sigma_8$ (b.f.) &$\chi^2\,({\rm PTE})$ \\
\hline
\textbf{All bins} (22) 
    & $0.25^{+0.02}_{-0.02}$ ($0.24$)
    & $0.82^{+0.07}_{-0.07}$ ($0.82$)
    & $0.91^{+0.10}_{-0.10}$ ($0.92$)
    & 31.2 (9\%) \\ 
\textbf{All bins + BAO} (23) 
    & $0.29^{+0.01}_{-0.01}$ ($0.29$)
    & $0.79^{+0.06}_{-0.06}$ ($0.75$)
    & $0.80^{+0.06}_{-0.06}$ ($0.76$)
    & 35.6 (4.5\%) \\
\textbf{No AAM} (22) 
    & $0.27^{+0.02}_{-0.02}$ ($0.25$)
    & $0.75^{+0.04}_{-0.04}$ ($0.80$)
    & $0.79^{+0.06}_{-0.06}$ ($0.90$)
    & 33.4 (6\%) \\
\textbf{No $p(z)$, $N_\ell$ marg.} (22) 
    & $0.25^{+0.02}_{-0.02}$ ($0.24$)
    & $0.82^{+0.06}_{-0.06}$ ($0.84$)
    & $0.91^{+0.09}_{-0.10}$ ($0.93$)
    & 32.0 (8\%) \\
\textbf{Linear HEFT} (13) 
    & $0.25^{+0.03}_{-0.03}$ ($0.25$)
    & $0.70^{+0.06}_{-0.06}$ ($0.70$)
    & $0.76^{+0.09}_{-0.08}$ ($0.77$)
    & 14.3 (36\%) \\ 
$k_{\rm max}= 0.25 \, \rm Mpc^{-1}$ (16)
    & $0.25^{+0.03}_{-0.03}$ ($0.25$)
    & $0.84^{+0.08}_{-0.08}$ ($0.82$)
    & $0.93^{+0.14}_{-0.13}$ ($0.90$)
    &  20.4 (20\%) \\ 
\textbf{Bin 1 + BAO} (3) 
    & $0.30^{+0.01}_{-0.01}$ ($0.30$)
    & $0.98^{+0.14}_{-0.13}$ ($1.10$)
    & $0.98^{+0.14}_{-0.13}$ ($1.10$)
    & 4.00 (26\%) \\
\textbf{Bin 2 + BAO} (8.5) 
    & $0.30^{+0.01}_{-0.01}$ ($0.30$)
    & $0.88^{+0.12}_{-0.12}$ ($0.94$)
    & $0.89^{+0.12}_{-0.12}$ ($0.95$)
    & 16.6 (4.4\%)\\
\textbf{Bin 3 + BAO} (10.5) 
    & $0.29^{+0.01}_{-0.01}$ ($0.29$)
    & $0.75^{+0.05}_{-0.05}$ ($0.73$)
    & $0.76^{+0.05}_{-0.05}$ ($0.74$)
    & 13.7 (22\%) \\
\hline
\end{tabular}
\caption[Constraints on $\Omega_m$, $S_8$ and $\sigma_8$ for different analysis configurations]{Constraints on $\Omega_m$, $S_8$ and $\sigma_8$ for different analysis configurations. In each case we quote the posterior mean and 68\% confidence interval, with the best-fit value in parentheses. These constraints can be visualised in Fig. \ref{fig:all_violins}. The $p$-value in each case was calculated by estimating the effective number of degrees of freedom as discussed in the text (shown in the first column).}
\label{tab:constraints_bestfit}
\end{table}

    These are compatible with the value preferred by \planck TT,TE,EE+lowE+lensing ($S_8=0.832\pm0.013$), as well as with most existing CMB lensing tomography analyses. Namely, Garc\'ia-Garc\'ia et al. 2021 \cite{2105.12108} find $S_8=0.825\pm0.023$ in the absence of cosmic shear data, using a suite of galaxy samples in the redshift range $0.25\lesssim z  \lesssim1.5$. Using quasars from the {\sl Quaia} sample, spanning the range $z\lesssim 3$, Piccirilli et al. 2024 \cite{2402.05761}  find $S_8=0.841\pm0.044$. Farren et al. 2024 \cite{2309.05659} in turn find $S_8=0.810\pm 0.015$ combining lensing from \planck and ACT with galaxies from the unWISE sample in the range $0.5\lesssim z\lesssim 1.5$. Finally, the analysis that most closely matches the redshift range explored here is that of Sailer et al. 2025 \cite{2503.24385}, combining the DESI Legacy photometric bright galaxy and luminous red galaxy samples (BGS and LRG respectively) with CMB lensing and lensing magnification. The combined constraint found there is $S_8=0.788\pm0.020$, while the measurement using only BGS at $z<0.4$ is $S_8=0.870^{+0.050}_{-0.064}$. Although our statistical uncertainties are larger than most of these past analyses, this study provides a valuable confirmation of these past constraints from a complementary range of redshifts ($z\lesssim0.3$).
    
  \begin{figure}
      \centering
      \includegraphics[width=1\linewidth]{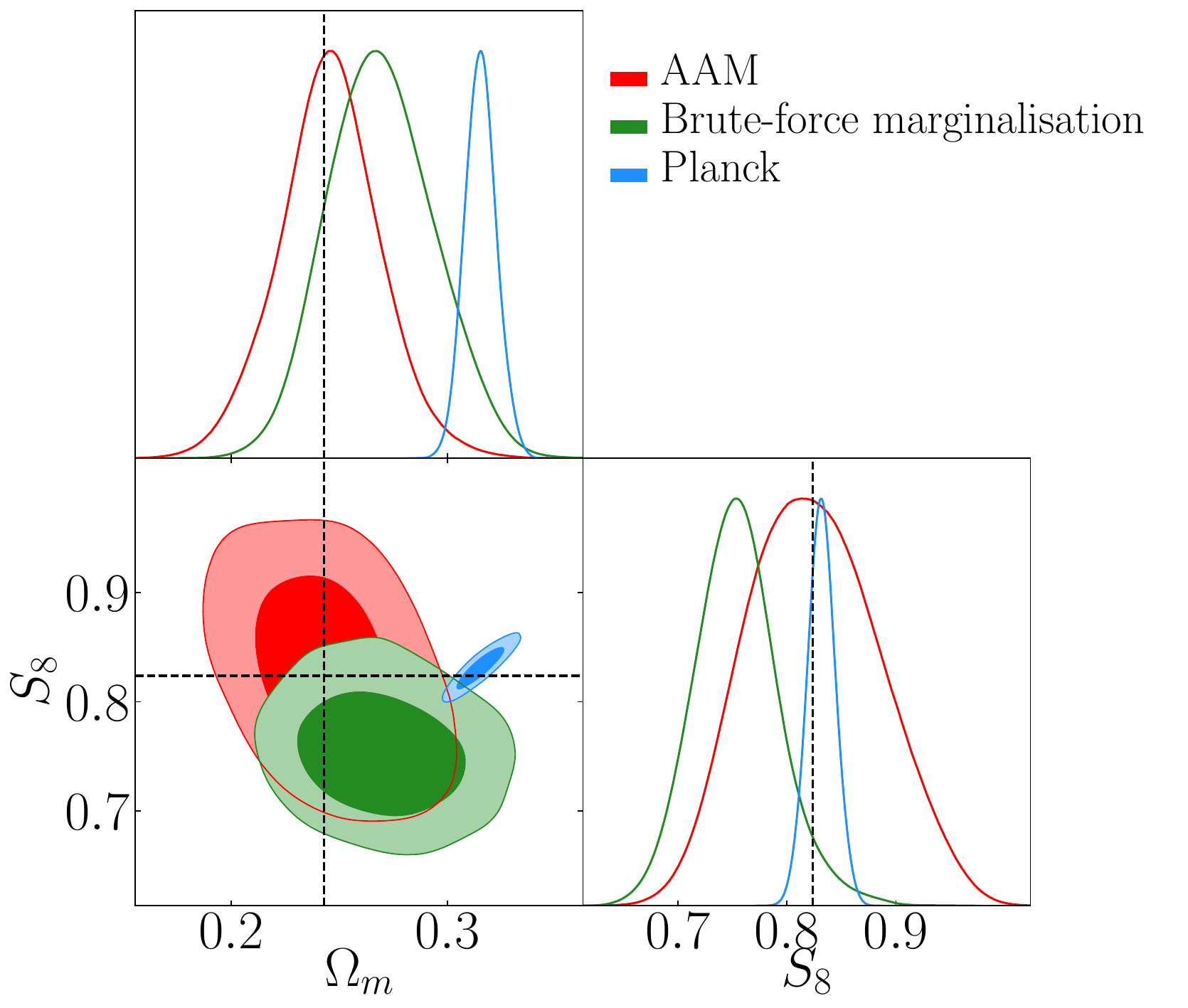}
      \caption[Posterior comparison between the AAM strategy over the bias factors and the \enquote{brute force} MCMC marginalisation]{
      Constraints on $S_8$ and $\Omega_m$ found in our fiducial analysis through the AAM strategy to marginalise over the HEFT bias parameters avoiding volume effects (red). These are compared against the result of brute-force marginalisation over the bias parameters (green), which results in a shift in the posterior contours due to volume effects. This can be seen by comparing the position of the posterior distribution with the best-fit model (marked by the dashed lines). The AAM technique avoids these volume effects by effectively imposing a Jeffreys-like prior.
      } \label{fig:all_bins_profile_vs_no_profile}
  \end{figure}

    Interestingly, we find that our data seems to prefer a low value of $\Omega_m$ when this parameter is left to vary freely: $\Omega_m=0.245\pm 0.024$, in tension with the \planck TT,TE,EE+lowE+lensing constraints at the $2.8\sigma$ level. In turn, the tension with the lower value of $\Omega_m$ preferred by the DESI measurements is significantly lower ($1.6\sigma$). As we show in Appendix \ref{app:om}, the constraint on $\Omega_m$ is dominated by the galaxy clustering auto-spectrum in the highest-redshift bin (Bin 3). This is because the constraint on $\Omega_m$ is based on the broadband shape of the power spectrum, which is measured with the highest sensitivity and over the broadest range of angular scales for this particular correlation. Given that a standard ruler measurement of $\Omega_m$ from BAO data is substantially more robust against potential systematic contamination in galaxy clustering than one based on the broadband shape of the spectrum, and since our constraints are broadly compatible with DESI, we will report most of our results regarding the amplitude of matter fluctuations and the growth of structure in combination with the DESI BAO prior in what follows. Importantly, this will also allow us to avoid potential systematic errors due to the extrapolation of the \bemu HEFT templates to low values of $\Omega_m$.

    The best-fit model has a $\chi^2$ of 32.1 with a free $\Omega_m$ and 35.5 when including the DESI prior. Our data vector has 34 elements and thus, in both cases, it is a likely realisation of both best-fit models. The goodness of fit of the models is less clear when accounting for the total number of degrees of freedom. Our model has 12 nuisance parameters (4 bias parameters per redshift bin) in addition to the two cosmological parameters. However, all model parameters enter the model non-linearly, and it is not clear whether $N_{\rm dof}=N_{\rm data}-N_{\rm param}=34-14=20$ is a good estimate of the effective number of degrees of freedom. To determine this, we generate 1000 Gaussian realisations of our data vector, taking the best-fit model found using the DESI prior as the mean, and the data covariance. We then find the best-fit model prediction for each realisation by minimising the $\chi^2$, and study the distribution of best-fit $\chi^2$ values. We find that it is reasonably well described by a $\chi^2$ distribution with an effective number of degrees of freedom $N_{\rm dof,eff}=22$. Using this distribution, we find that the probability-to-exceed (PTE, or $p$-value) of the $\chi^2$s found in the data is 0.09 in the free $\Omega_m$ case, and 0.045 when using the DESI BAO prior. Although both values are relatively low, they are not entirely unlikely (corresponding to $\sim 1.7\sigma$ and $\sim2\sigma$ fluctuations, respectively).  Therefore, and given that the data itself is well described by these models, we consider the goodness of fit acceptable. When calculating the PTE for any other data combination used here, we follow the procedure described above to determine the effective number of degrees of freedom. 
    Our results so far have been obtained by marginalising over the HEFT bias parameters using AAM (see Section \ref{sssec:method.like.prof} and Ref. \cite{2301.11978}), in order to avoid volume effects due to marginalisation over a large nuisance parameter space. To determine the impact of these effects in our analysis, we recalculate our constraints using \enquote{brute-force} marginalisation over the bias parameters (i.e. including them as free parameters in the MCMC chain). The results are shown in Fig. \ref{fig:all_bins_profile_vs_no_profile}, which also displays the best-fit model. We find that volume/projection effects are in fact sizeable, disfavouring the larger values of $S_8$, shifting the marginalised posterior distribution for $S_8$ downwards by more than $1\sigma$ with respect to the best fit, and leading to an artificial shrinkage of the final uncertainties. In turn (and by construction), the posterior distribution using analytical marginalisation is centred at the best-fit and, since we are effectively using a Jeffreys prior on the bias parameters, the width of this distribution is not affected by volume effects. 
    
    To validate our results against the impact of uncertainties in the modelling of non-linear bias we carry out two further tests. First, we repeat our fiducial analysis, without the DESI BAO prior, using a more conservative scale cut, $k_{\rm max}=0.25\,{\rm Mpc}^{-1}$. The result of this test is shown in Fig. \ref{fig:all_violins}. We find that the uncertainties on both $\Omega_m$ and $S_8$ increase (by 50\% and 14.5\%, respectively), but their preferred values are not shifted significantly from the constraints found in our fiducial analysis. Secondly, we repeat our analysis using a simpler bias model, setting $b_2=b_s=b_{\nabla^2}=0$, leaving only $b_1$ as a free parameter, and restricting the range of scales to $k<0.11\,{\rm Mpc}^{-1}$. We refer to this case as ``linear HEFT''. For comparison, this is similar to the ``model independent'' setup of \cite{2407.04607}\footnote{\cite{2407.04607} also consider a linear bias model, although in that case they marginalise over counterterm parameters to account for residual systematic shifts due to non-linearities.} In this case we find a $\sim1.1\sigma$ downward shift in $\sigma_8$. Interestingly a similar shift is found in \cite{2407.04607} in their model-independent analysis. This suggests that the impact of non-linear and scale-independent bias can extend to relatively large scales. Interestingly, we find that the value of $\Omega_m$ preferred by the data is not affected by the linear bias assumption. This reinforces the results described in Appendix \ref{app:om}, which show that the constraints on this parameter are dominated by the large-scale broadband shape of the matter power spectrum in the highest redshift bin (Bin 3).

    To test that our results are not significantly affected by large-scale systematic contamination in the observed galaxy number counts, we repeated our analysis using a more conservative large-scale cut of $\ell_{\rm min}=40$. This results in only a small $\sim0.3\sigma$ upwards shift in the best-fit value of $S_8$, accompanied by a $\sim20\%$ increase in its statistical uncertainties. We also tested that marginalising over the potential contamination of zero-point variations in the zero-point of SuperCOSMOS plates, as described in \cite{1909.09102}, does not affect our constraints significantly, leading to similarly small shifts of $\lesssim0.3\sigma$ in $S_8$.

    \begin{figure}
      \centering
      \includegraphics[width=1\linewidth]{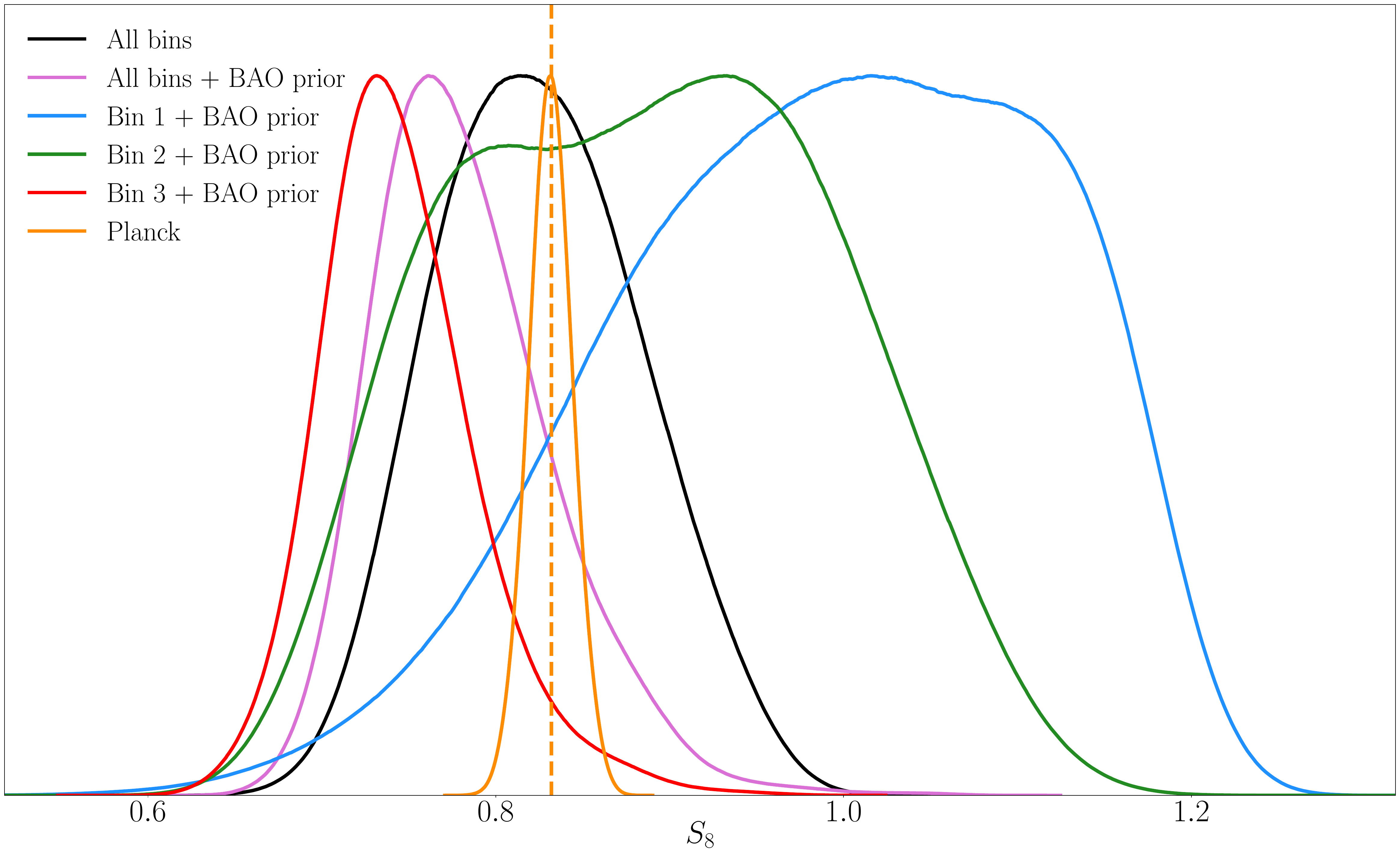}
      \caption{Marginalised posterior distribution on $S_8$ obtained from the combination of all redshift bins leaving $\Omega_m$ free (black). When assuming a BAO prior on $\Omega_m$ from DESI, we obtain the marginalised posterior shown in pink for the joint fit of the three redshift bins, and the results for the individual redshift bins are shown in blue, green, and red for bins 1, 2, and 3, respectively. For comparison, we show also the \planck constraints (orange). As can be seen, the constraining power is dominated by Bin 3, with the information in bins 1 and 2 limited by the range scale used in the analysis, and the lower signal-to-noise ratio of the CMB lensing cross-correlation there.}\label{fig:s8_bins}
    \end{figure}

  \subsection{Low-redshift tomographic growth reconstruction}\label{ssec:results.s8z}
     Having measured the galaxy auto-correlation and cross-correlation with CMB lensing in three different redshift bins allows us, in principle, to recover the amplitude of matter fluctuations as a function of redshift, and thus reconstruct the growth history at late times. As we noted in the previous section, and shown in Appendix \ref{app:om}, our constraints on $\Omega_m$ are driven by the galaxy auto-correlation in the last redshift bin and, unfortunately, the reduced range of angular scales over which the lower-redshift correlations can be used, does not allow us to simultaneously measure both $\sigma_8$ and $\Omega_m$ in the first two redshift bins. As we did in the previous section, we will therefore resort to constraining $\Omega_m$ by imposing the DESI BAO prior when obtaining per-bin constraints. 

     Fig. \ref{fig:s8_bins} shows the constraints on $S_8$ found in our fiducial setup, both with and without the BAO prior, and the constraints obtained from each redshift bin independently. These results are also listed in Table \ref{tab:constraints_bestfit}. We find that, as expected, our joint constraints are dominated by the measurement in the highest redshift bin. The constraints obtained from the two lower-redshift bins are significantly broader and compatible with our fiducial measurement, although they also allow for larger values of $S_8$. In all cases, the constraints found are in good agreement with the \planck measurements, also shown in the figure.
\begin{figure}
    \centering
    \includegraphics[width=0.9\linewidth]{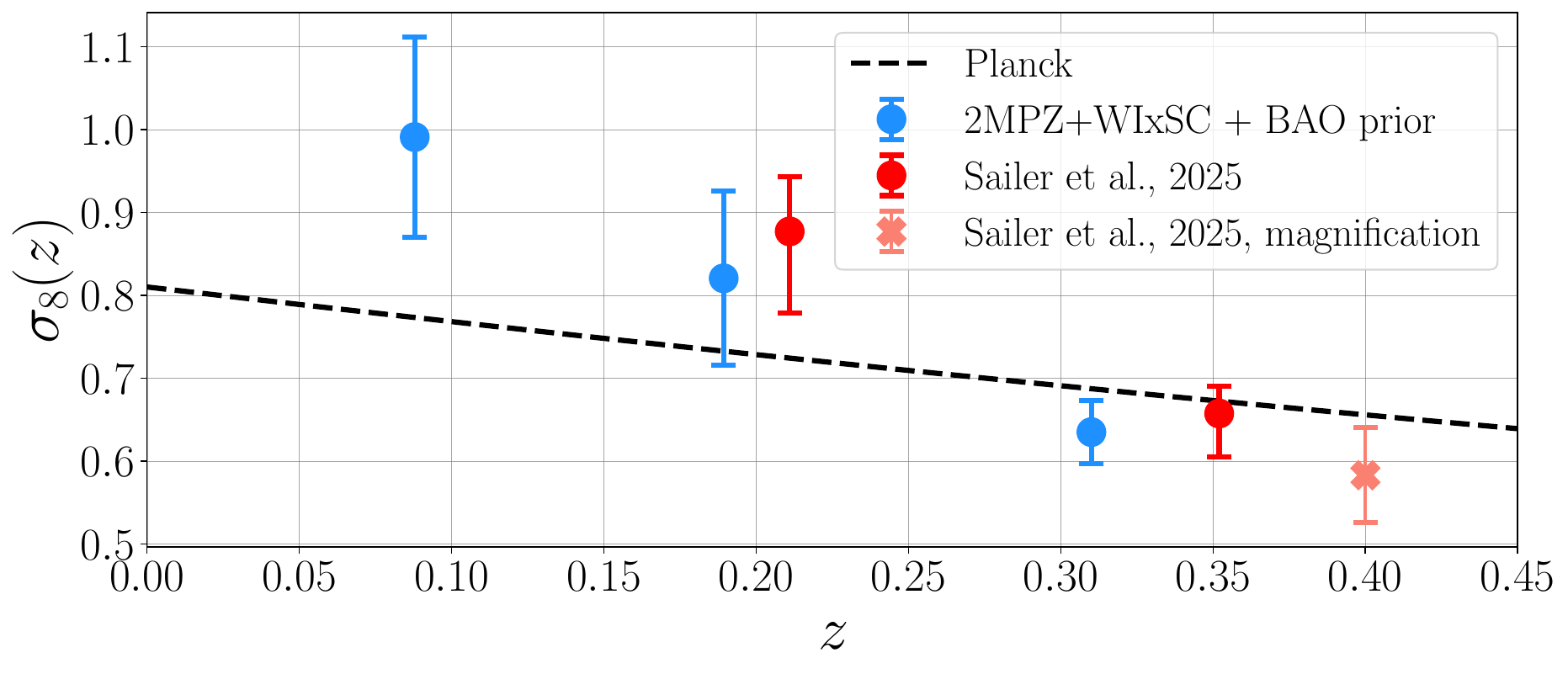}
    \caption[Growth reconstruction based on the bin-by-bin measurement of $\sigma_8$]{Growth history (in terms of $\sigma_8(z)$) reconstructed from our measurements of $C_\ell^{gg}$ and $C_\ell^{g\kappa}$ in each redshift bin, assuming a BAO prior on $\Omega_m$ from DESI (blue). Our results are in reasonable agreement with \planck (dashed line), and with previous low-redshift measurements from \cite{2503.24385} (red and pink).}
    \label{fig:growth_reconstruction}
\end{figure}

     In Fig. \ref{fig:growth_reconstruction} we present these constraints in terms of the redshift-dependent $\sigma_8(z)$, defined as $\sigma_8(z)=\sigma_8\,D(z)$, where $D(z)$ is the linear large-scale growth factor, normalised to $D(0)=1$. We calculate $\sigma_8(z)$ as a derived parameter in terms of both $\sigma_8$ and $\Omega_m$, since $D(z)$ depends on the latter. Our numerical constraints are:
     \begin{equation}\nonumber
       \sigma_8(z=0.074)=0.99\pm0.12,\hspace{12pt}
       \sigma_8(z=0.18)=0.82\pm0.11,\hspace{12pt}
       \sigma_8(z=0.30)=0.635\pm0.038.
     \end{equation}
     The recovered growth history is in relatively good agreement with the \planck prediction, although a marginally steeper growth is preferred, with the lowest- and highest-redshift measurements lying $1.8\sigma$ above and $1.4\sigma$ below the \planck values, respectively. Our measurements are also in relatively good agreement with the constraints of \cite{2503.24385}, to our knowledge the only other lensing tomography analysis targetting the low-redshift regime using HEFT. Interestingly, the constraints from \cite{2503.24385} also show a trend towards a marginally steeper growth at low redshifts, although compatible with the \planck prediction within uncertainties. These results are also in generally good agreement with other low-redshift tomographic analyses. Using \tmpz, \cite{1801.03736} found that the amplitude of fluctuations at late times is broadly compatible with the \planck prediction. A similar result was found by \cite{1805.11525} combining \tmpz and \wisc, also finding a trend towards large $\sigma_8$ at the lowest redshifts. Compatible results were also found by \cite{1908.04854}. We must note, however, the substantial differences in the bias models and analysis choices used in the latter works, making a direct comparison with our results difficult in detail.

  \subsection{Constraints on bias parameters and coevolution relations}\label{ssec:results.bias}
    \begin{figure}
        \centering
        \includegraphics[width=0.8\linewidth]{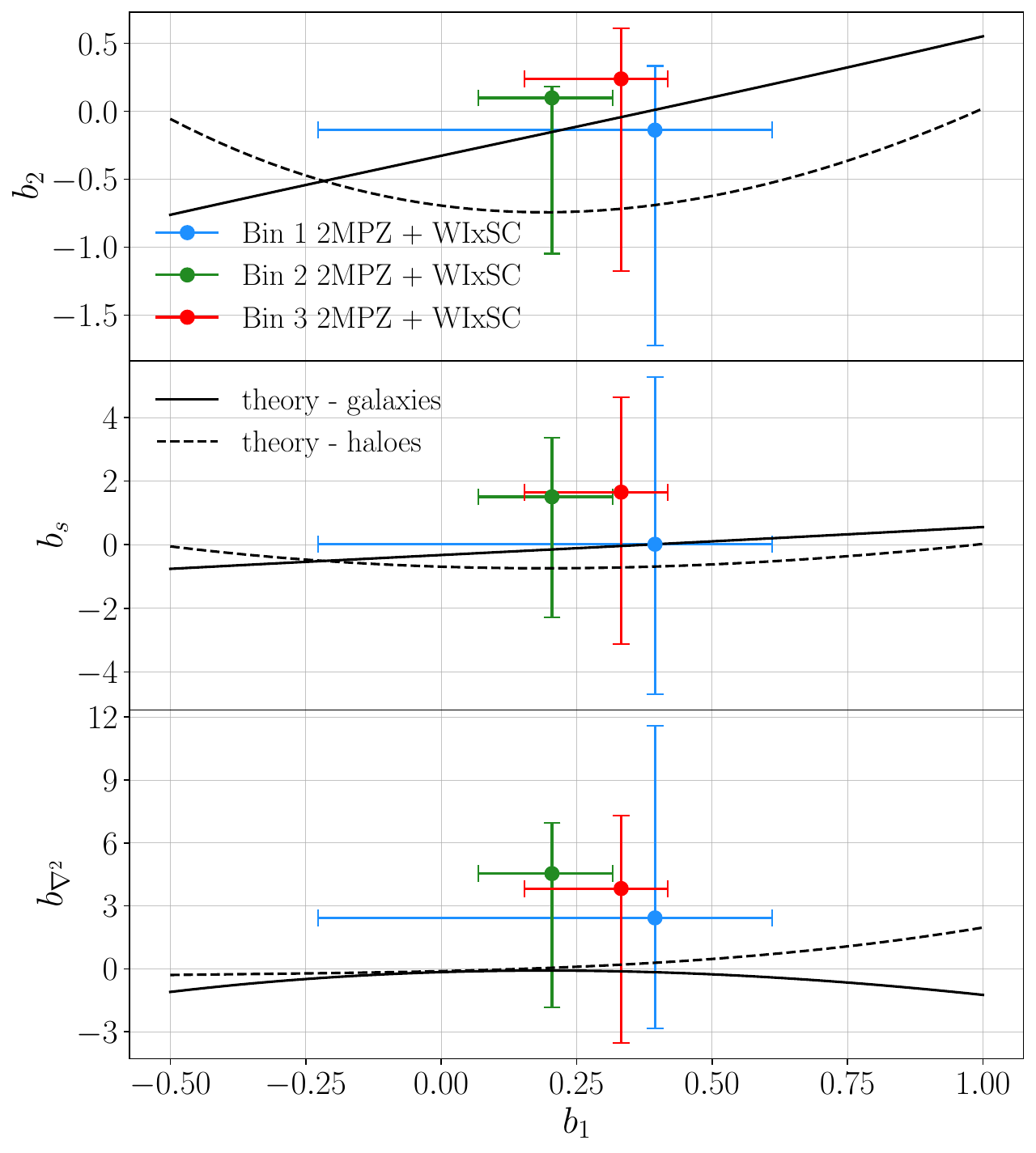}
        \caption[Coevolution relations]{Constraints on the higher-order HEFT bias parameters ($b_2$, $b_s$, $b_{\nabla^2}$) as a function of the linear Lagrangian bias $b_1$. Our constraints are compared with the coevolution relations of \cite{2110.05408} for haloes and galaxies (dashed and solid lines, respectively). Note that, in contrast with Bins 2 and 3, the Bin 1 galaxy sample contains galaxies from 2MASS. This results in different galaxy populations and, whereas $b_1$ grows with redshift, as expected for Bins 2 and 3, Bin 1 has the largest value.}
        \label{fig:coevolution_vs_b1}
    \end{figure}

\begin{table}
    \centering
    \renewcommand{\arraystretch}{1.5}
    \begin{tabular}{rl}
        \hline
        \multicolumn{2}{c}{Haloes, $k_{\rm d} = 0.75 \ihMpc$, $k_{\rm max} = 0.7 \ihMpc$}\\
        \hline
        $b_{2}\left(b_1\right)=$ & $2 \times \left [-0.09143 \left ( b_1 \right )^3 + 0.7093 \left(b_1\right)^2 - 0.2607 b_1- 0.3469 \right]$\\
        $b_{s}\left(b_1 \right)=$ & $2 \times \left [ \left(0.02278 (b_1\right)^3 - 0.005503 \left(b_1\right)^2 - 0.5904 b_1 - 0.1174 \right]$\\
        $b_{\nabla^2}\left(b_1\right)=$ & $-0.6971 \left(b_1 \right)^3 + 0.7892 \left(b_1\right)^2 + 0.5882 b_1 - 0.1072$\\
        \hline
        \hline
        \multicolumn{2}{c}{Galaxies, $k_{\rm d} = 0.75 \ihMpc$, $k_{\rm max} = 0.7 \ihMpc$}\\
        
        $b_{2}\left(b_1\right)=$ & $2 \times \left[0.01677 \left(b_1 \right)^3 - 0.005116 \left(b_1\right)^2 + 0.4279 b_1 - 0.1635 \right]$\\
        $b_{s}\left(b_1\right)=$ & $2 \times \left[-0.3605 \left(b_1\right)^3 + 0.5649 \left(b_1\right)^2 - 0.1412 b_1 - 0.01318 \right]$\\
        $b_{\nabla^2} \left(b_1 \right)=$ & $0.2298 \left(b_1 \right)^3 - 2.096 \left(b_1 \right)^2 + 0.7816 b_1- 0.1545$\\
        \hline
    \end{tabular}
    \caption{The phenomenological fitting functions for the bias coevolution relations obtained by \cite{2110.05408} for haloes and galaxies, in the case with  smoothing $k_{\rm d} = 0.75 \, \ihMpc$ and $k_{\rm max} = 0.7 \, \ihMpc$. Note the different prefactors accounting for the different definition of $b_2$ and $b_s$ between our Eq. \ref{eq::advection} and \cite{2110.05408}.}
    \label{tab:polyfits}
\end{table}

    \noindent While extracting cosmological information and reconstructing the growth function at low redshift are our primary goals, analysing the values assumed by the galaxy bias parameters is also interesting. In particular, higher-order bias parameters have been found to exhibit correlations with the corresponding value of linear bias, often called ``coevolution relations'' \citep{1511.01096,1611.09787,1712.07531,1904.11294,2110.05408,2412.06886}. Constraining these correlations, albeit difficult, might help reduce the parameter space to be explored in cosmological analyses. We also note that, for analyses including scales $k > 0.1$-$0.2 ~ \mathrm{Mpc}^{-1}$, the effect of baryons on the cross-correlations between matter and galaxy fields should be accounted for \cite{2412.08623}. However, given the error bars of our data set and the fact that we do not include observables that probe extremely small scales (such as $k=5$ or $10 ~ \mathrm{Mpc}^{-1}$), we find that any baryonic model would be unconstrained and we therefore choose to not consider the effect of baryons in this case.
    
    Fig. \ref{fig:coevolution_vs_b1} displays the relations between the higher order bias parameters ($b_2$, $b_s$ and $b_{\nabla^2}$) and the linear bias $b_1$, as inferred in our analysis. Like in our fiducial analysis, we jointly fit the data vectors of our three redshift bins. However, in this case, since we are interested in quoting inferred marginalised posterior distributions of the bias parameters, we necessarily avoid using the AAM method and explicitly sample the galaxy bias parameters (what we called before ``brute force marginalisation''). This results in poorly converged chains when both cosmology and nuisance parameters are left free. For this reason, we present these results assuming a \planck cosmology, which we keep fixed while sampling the galaxy bias parameters. We have checked that the qualitative results remain the same for two different cosmologies, with higher and smaller $\sigma_8$.

    We compare our inferred parameters to fitting functions of the coevolution relations present in the literature. Although different works have explored these relations in the context of Eulerian bias expansions both for haloes \citep[e.g.][]{1511.01096,1712.07531} and for galaxies \citep{2105.02876}, these relations are not easily translated to a Lagrangian framework. More recently \cite{2110.05408,2412.06886} have shown different fitting functions for haloes and various types of galaxies in the context of hybrid Lagrangian bias expansions. Specifically, we will use the fitting functions obtained in \cite{2110.05408} using the same HEFT implementation as in this work, with the same smoothing of the initial Lagrangian fields. We modify these fitting functions to account for the different numerical prefactors between the expansion assumed in Eq. \ref{eq::advection} and in \cite{2110.05408}, and show the modified functions in Table \ref{tab:polyfits}.
     
    At the linear bias level, we find that $b_{1,1} > b_{1,3} > b_{1,2}$. The relation between bins 2 and 3 is the one expected: galaxy bias tends to grow with redshift for samples with a similar apparent magnitude limit. The reason why $b_{1,1}$ is larger than the others, despite Bin 1 being the lowest redshift sample, is that it probes a different galaxy population. The \tmpz sources contained in Bin 1 are selected by cross-matching 2MASS in addition to WISE and SuperCOSMOS, whereas bins 2 and 3 were selected from the \wisc survey, constructed using only the latter two catalogues. 

    When considering the relations between higher-order bias parameters and the corresponding linear bias, we compare against both the coevolution relation calibrated on galaxies (black-solid lines in Fig. \ref{fig:coevolution_vs_b1}) and on Dark Matter haloes (black-dashed lines). We find that our inferred points lie within $1\sigma$ of the theoretical predictions. Some level of scatter is expected, since these coevolution relations strongly depend on the galaxy sample. Specifically, galaxy assembly bias and selection effects can amplify the scatter around these phenomenological fitting functions. These results are in agreement with \cite{2407.07949}, where the authors also used HEFT to constrain galaxy bias parameters (in this case, using 3D clustering from the SDSS-III BOSS survey) and found them to be scattered around -- but overall compatible with -- these coevolution relations. 
 
\section{Conclusions}\label{sec:conc}

  We set about measuring the growth of structure at late times, using the combined galaxy clustering of photometric samples from the \tmpz and \wisc surveys and their cross-correlation with the CMB lensing signal as measured by \planck. Constraining the evolution in the amplitude of matter fluctuations at late times ($z\lesssim0.3$) is important, as it covers the dark energy-dominated epoch, and constitutes a range of redshifts that is complementary to that explored by recent lensing tomography studies \cite{2105.12108,2407.04607,2309.05659,2402.05761,2503.24385}.

  Combining our measurements with a prior on $\Omega_m$ based on the BAO measurements of \cite{2503.14738}, we find $S_8=0.79\pm0.06$, or $\sigma_8=0.80\pm0.06$, both in agreement with \planck as well as most lensing tomography analyses. Interestingly, we find that, when dropping this BAO prior, our measurements favour a low value of $\Omega_m=0.245\pm0.024$,  in tension with the \planck measurements at the $2.8\sigma$ level. Other projected large-scale structure analyses at intermediate redshifts have found evidence of a low $\Omega_m$ at varying significance, including \cite{2403.13794} using cosmic shear, and \cite{2010.00466,2309.05659,2410.10808} using CMB lensing tomography. Other analyses using data at similar redshifts (e.g. \cite{2407.04607,2507.07968}) have found no such evidence, however. As we show in Appendix \ref{app:om}, the information on $\Omega_m$ is dominated by the broadband shape of the galaxy auto-correlation (particularly from the higher redshift bin, in our case). This constraint is thus significantly less robust than the BAO standard ruler measurement against potential clustering systematics, as well as misspecification of the non-linear bias model. Both would affect the clustering of various types of galaxies in different ways, which could explain the differences between these analyses (e.g. \cite{2407.04607,2507.07968} employed luminous red galaxies, whereas other works, including the present one used other samples).

  Analysing each redshift bin independently, we have also reconstructed the growth of density fluctuations at late times, which we find to be in agreement with the growth predicted by \planck. Our results show some hints of a preference for a steeper growth at late times, 
  with similarly weak evidence present in previous analyses \cite{2105.12108,2503.24385}. Although the statistical uncertainties are too large to derive any definite conclusions from these constraints, it is interesting to consider when lensing tomography may be able to provide sufficiently precise measurements of $\sigma_8(z)$ to place meaningful constraints on the physics of dark energy. This is particularly relevant in the context of the recent evidence for dynamical dark energy from the combination of CMB, BAO, and supernova data \cite{2504.06118}. Observing a similar departure from a cosmological constant from growth probes would be vital to confirm this evidence as new physics, and to ascertain the nature of the components driving the late-time accelerated expansion. Although redshift-space distortion measurements may be able to provide this evidence with future data from DESI and Euclid \cite{1110.3193}, lensing tomography measurements from near-future photometric imaging surveys, such as the Rubin Observatory \cite{0805.2366, 1809.01669}, and ground-based CMB experiments, such as the Simons Observatory \cite{1808.07445}, could also reach the precision needed to address this question \cite{1808.07445}. Further progress can be made with existing data, however. In particular, replacing CMB lensing with cosmic shear measurements from current surveys, such as DES, KiDS, and HSC \cite{2105.13544,2304.00701,2503.19441}, could provide more sensitive measurements of the galaxy-lensing cross-correlations for low-redshift galaxy samples, allowing us to improve the bounds on $\sigma_8(z)$ in the $z\lesssim0.3$ range. Such effort is all the more timely given the most recent results from the HSC, DES and DESI collaborations \citep{2026arXiv260114559D,2025arXiv251118134C,2025arXiv251215962L}.

  At the low redshifts probed here, it becomes imperative to use a galaxy bias model that can accurately describe the clustering of galaxies and matter down to relatively small scales, since these scales inevitably contribute to the measured angular statistics. However, it is interesting to consider the extent to which useful cosmological information can actually be extracted from galaxy clustering on non-linear scales. In the case of projected clustering (i.e. in the absence of redshift-space distortions), most past analyses have found that the additional model complexity required to describe the galaxy power spectrum on mildly non-linear scale approximately offsets the information gained from the additional modes unlocked \cite{2307.03226,2309.05659,2407.04607}. Similar results have been found for simpler ``physics-agnostic'' parametrisations \cite{2508.05319}. In our case, for example, the uncertainties on $\sigma_8$ obtained in the ``linear HEFT'' bias model for $k<0.11\,{\rm Mpc}^{-1}$ are in fact $\sim10\%$ smaller than those found in our fiducial analysis for $k<0.42\,{\rm Mpc}^{-1}$, in spite of the $\sim4$-fold increase in $k_{\rm max}$ allowed by the use of the full HEFT parametrisation. Further gains may be made by including information from higher-order statistics, such as the bispectrum, which can be used to self-calibrate the bias parameters, allowing us to recover more information from the small-scale power spectrum. Although current studies including the projected galaxy bispectrum have only found a relatively mild reduction ($\sim10$-$20\%$) in the final parameter errors \cite{2507.07968,Verdiani_251017796}, larger improvements may be achieved with future, more sensitive data, as well as more sophisticated bias models able to simultaneously predict the galaxy power spectrum and bispectrum on small scales.
  
\acknowledgments
  We would like to thank Noah Sailer for useful comments on this manuscript. 

  AR acknowledges that this publication was produced while attending the PhD programme in Space Science and Technology at the University of Trento  (\url{https://phd.unitn.it/phd-sst/}), Cycle XXXVIII, with the support of a scholarship financed by the Ministerial Decree no. 351 of 9th April 2022, based on the NRRP - funded by the European Union - NextGenerationEU - Mission 4 \enquote{Education and Research}, Component 1 \enquote{Enhancement of the offer of educational services: from nurseries to universities} - Investment 4.1 \enquote{Extension of the number of research doctorates and innovative doctorates for public administration an
  d cultural heritage} [CUP E63C22001340001].  
  AR acknowledges support from the Research grant TAsP (Theoretical Astroparticle Physics) funded by INFN. 

  CGG, DA, and MZ acknowledge support from the Beecroft Trust. We made extensive use of computational resources at the University of Oxford Department of Physics, funded by the John Fell Oxford University Press Research Fund. REA received support from grant PID2024-161003NB-I00 funded by
MICIU/AEI/10.13039/501100011033 and by ERDF/EU.

\appendix
\section{Sensitivity to $\Omega_m$}\label{app:om}
  In order to identify the part of our data vector that governs the observed preference for a low value of $\Omega_m$, we reanalyse the data in a more model-agnostic manner, isolating only the shape information present in the measurements. Specifically, we turn to the empirical bias model presented in \cite{2508.05319}. In this model, the galaxy-galaxy and galaxy-matter power spectra are modelled as
  \begin{equation}\label{eq:bias_maleubre}
    P_{gg}=(A_{gg}^0+k^2A_{gg}^2)P_{mm}(k)+N_{gg}^0,\hspace{12pt}
    P_{gm}=(A_{gm}^0+k^2A_{gm}^2)P_{mm}(k)+N_{gm}^0.
  \end{equation}
  Here, $\{A_{xy}^n,N_{xy}^n\}$ are free parameters, and $P_{mm}(k)$ is the matter power spectrum. The white noise parameters $N_{xy}^0$ absorb the impact of shot noise and any stochastic bias components, both in $P_{gg}$ and $P_{gm}$. In turn, the amplitude parameters $A_{xy}^0$ and $A_{xy}^2$ account for linear and scale-dependent bias, respectively. This model has been shown to provide an accurate description of the clustering of galaxies and their correlation with the matter overdensity up to scales $k_{\rm max}=0.3\,{\rm Mpc}^{-1}$. Furthermore, since we do not relate $A_{gg}^0$ and $A_{gm}^0$ in terms of the linear galaxy bias, these parameters are fully degenerate with any cosmological parameter controlling the amplitude of the matter power spectrum, namely $\sigma_8$. This allows us to use this model to isolate the information contained in the shape of the power spectrum.

  We use this model to derive constraints on $\Omega_m$ from different sectors of our data vector, marginalising over all free amplitude and noise parameters. As in the main analysis, to avoid volume effects we use the AAM technique to perform this marginalisation. In this case, since all nuisance parameters are linear, the analytical marginalisation is exact. We assign different nuisance parameters $\{A_{xy}^n,N_{xy}^0\}$ to each redshift bin, use the same large-scale cuts employed in our main analysis, and impose a small-scale cut $k_{\rm max}=0.3\,{\rm Mpc}^{-1}$, where this model has been shown to be sufficiently precise.

  \begin{figure}
    \centering
    \includegraphics[width=0.49\linewidth]{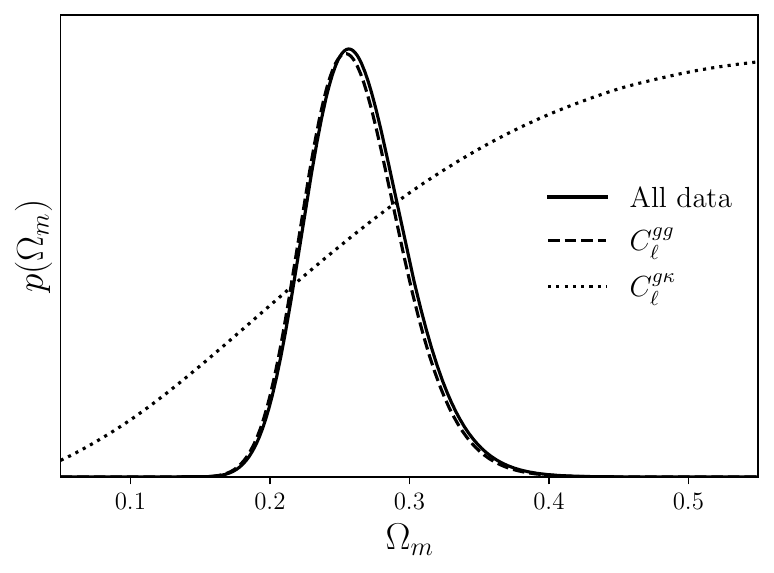}
    \includegraphics[width=0.49\linewidth]{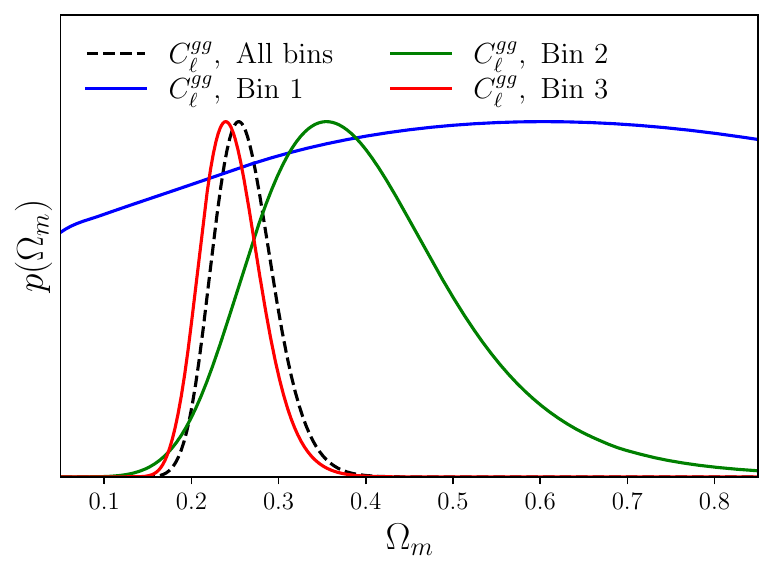}
    \caption{Constraints on $\Omega_m$ found using the empirical bias model of Eq. \ref{eq:bias_maleubre}. {\sl Left:} constraints found from the full data vector (solid), using only galaxy auto-correlations (dashed), and using only galaxy-$\kappa$ cross-correlations (dotted). {\sl Right:} constraints found from the full set of galaxy auto-correlations (dashed black), and from the auto-correlations in each individual redshift bin (blue, green and red for bins 1, 2, and 3, respectively). The constraints on $\Omega_m$ are dominated by the shape of the power spectrum measured in the galaxy auto-correlation of the third redshift bin.}
    \label{fig:om_shape}
  \end{figure}
  
  First, to ascertain how much information is present in $C_\ell^{gg}$ and $C_\ell^{g\kappa}$, we derive constraints from our full data vector and from either correlation type separately. The result, displayed in the left panel of Fig. \ref{fig:om_shape}, shows that $C_\ell^{gg}$ dominates the constraint on $\Omega_m$. This makes sense, as $C_\ell^{gg}$ is measured at much higher significance than $C_\ell^{g\kappa}$ (see Fig. \ref{fig:cl_plot_wisc_all}). The right panel of Fig. \ref{fig:om_shape} then shows the constraints derived from the galaxy auto-correlation of each individual redshift bin, in comparison with the combined constraint. The constraints are clearly dominated by the highest-redshift bin, given our ability to measure the power spectrum over a wider range of angular scales. In contrast, the lowest redshift bin is largely insensitive to $\Omega_m$, and Bin 2 is only able to place a loose constraint on it by comparison.
  
  \begin{table}
    \centering
    \begin{tabular}{|l|c|c|c|c|}
      \hline
      \textbf{Configuration} & $\Omega_m$ & $\chi^2$ & $N_{\rm dof}$ & PTE\\
      \hline
      \textbf{All bins}, $gg+g\kappa$ & $0.264\pm 0.035$ & 9.00 & 11 & 62\% \\
      \textbf{All bins}, $gg$         & $0.262\pm 0.035$ & 4.59 & 5 & 47\% \\
      \textbf{Bin 2}, $gg+g\kappa$    & $0.41\pm 0.12$ & 1.81 & 3 & 61\% \\
      \textbf{Bin 3}, $gg+g\kappa$    & $0.250\pm 0.036$ & 5.28 & 5 & 38\% \\
      \textbf{Bin 2}, $gg$            & $0.40\pm 0.12$ & 0.7 & 1 & 41\% \\
      \textbf{Bin 3}, $gg$            & $0.248\pm 0.036$ & 2.65 & 2 & 27\% \\
      \hline
    \end{tabular}
    \caption{Constraints on $\Omega_m$ found using the empirical bias model of Eq. \ref{eq:bias_maleubre} for the different data configurations explored in Appendix \ref{app:om}. We omit all configurations unable to place meaningful constraints on $\Omega_m$ (e.g. those using $C_\ell^{g\kappa}$ alone or the first redshift bin).}
    \label{tab:constraints_om_shape}
  \end{table}

  The results of this reanalysis are shown in Table \ref{tab:constraints_om_shape}. Note that we obtain good fits to the data in all cases, with PTE values in the range 27\%-62\%. Since, in this case, all model parameters are linear, we can safely calculate the number of degrees of freedom as $N_{\rm dof}=N_{\rm data}-N_{\rm param}$.

\section{Full model constraints}\label{app:full}
\begin{figure}
  \centering
  \includegraphics[width=1\linewidth]{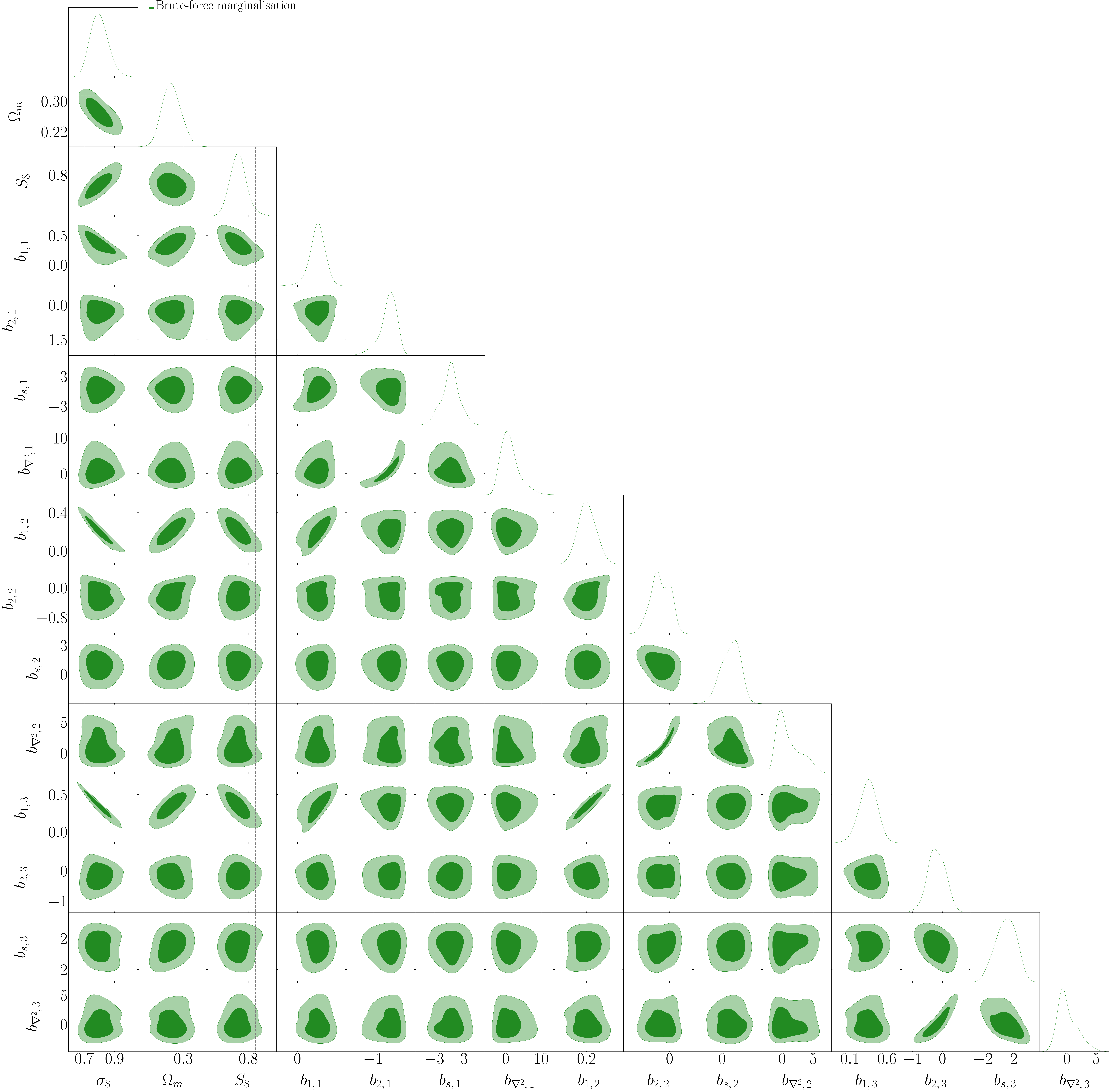}
  \caption{Constraints on all model parameters found via brute-force marginalisation. The gray dashed line shows Planck 2018 best fit \cite{Planck_2020}.}
  \label{fig:all_bins_profile_vs_no_profile_all_params}
\end{figure}
Figure \ref{fig:all_bins_profile_vs_no_profile_all_params} displays the constraints on the full set of bias and cosmological parameters entering the model when using brute-force marginalisation.

\newpage
\newpage
\bibliography{biblio}{}
\end{document}